\documentclass[twocolumn,epjc3]{svjour3} % use the EPJC template, two columns
\smartqed  % flush right qed marks, e.g. at end of proof
 \usepackage{amsmath}
 \usepackage{amssymb}
 \usepackage{graphicx}
  \usepackage[merge,numbers,compress]{natbib}
 \usepackage[T1]{fontenc}
 \usepackage{booktabs}
 \usepackage{xcolor}
 \usepackage{xspace}
 \usepackage{dcolumn}
 \usepackage{hyperref}
 \usepackage{caption}
 \usepackage{tabularx}
 \usepackage{lmodern}
 \usepackage{url}
 \usepackage{morefloats}
 \usepackage[utf8]{inputenc}
 \usepackage{hyphenat}
\let\oldhref\href
\renewcommand{\href}[2]{\oldhref{#1}{\hbox{#2}}}

\newcommand{\Pl}{\ell}

\def\be{\begin{equation}}
\def\ee{\end{equation}}

\newcommand{\PH}{\ensuremath{\text{H}}\xspace}
\newcommand{\Pj}{\ensuremath{\text{j}}\xspace}
\newcommand{\Pp}{\ensuremath{\text{p}}\xspace}
\newcommand{\Pe}{\ensuremath{\text{e}}\xspace}

\newcommand{\Pt}{\ensuremath{\text{t}}\xspace}

\newcommand{\PW}{\ensuremath{\text{W}}\xspace}
\newcommand{\PZ}{\ensuremath{\text{Z}}\xspace}

\newcommand{\Mt}{\ensuremath{m_\Pt}\xspace}

\newcommand{\MWOS}{\ensuremath{M_\PW^\text{OS}}\xspace}

\newcommand{\MZOS}{\ensuremath{M_\PZ^\text{OS}}\xspace}
\newcommand{\MZ}{\ensuremath{M_\PZ}\xspace}

\newcommand{\Gt}{\ensuremath{\Gamma_\Pt}\xspace}
\newcommand{\GH}{\ensuremath{\Gamma_\PH}\xspace}

\newcommand{\GZOS}{\ensuremath{\Gamma_\PZ^\text{OS}}\xspace}

\newcommand{\GWOS}{\ensuremath{\Gamma_\PW^\text{OS}}\xspace}

\newcommand{\GeV}{\ensuremath{\,\text{GeV}}\xspace}
\newcommand{\TeV}{\ensuremath{\,\text{TeV}}\xspace}

\newcommand{\alphas}{\ensuremath{\alpha_\text{s}}\xspace}

\newcommand{\ptsub}[1]{\ensuremath{p_{\text{T},#1}}\xspace}

\newcommand{\MVOS}{\ensuremath{M_{V}^\text{OS}}\xspace}%
\newcommand{\GVOS}{\ensuremath{\Gamma_{V}^\text{OS}}\xspace}%

\newcommand{\newc}{\newcommand}
\newc{\bi}{\begin{itemize}}
\newc{\ei}{\end{itemize}}
\newc{\benu}{\begin{enumerate}}
\newc{\eenu}{\end{enumerate}}
\newc{\bc}{\begin{center}}
\newc{\ec}{\end{center}}
\newc{\bfig}{\begin{figure}}
\newc{\efig}{\end{figure}}
\newc{\qbar}{\bar{q}}
\newc{\go}{\tilde{g}}
\newc{\PB}{\textsc{Powheg-Box}}

\newcolumntype{.}{D{.}{.}{-1}}
\newcolumntype{d}[1]{D{.}{.}{#1}}

\colorlet{tableoverheadcolor}{gray!37.5}
\colorlet{tableheadcolor}{gray!25}
\colorlet{tablerowcolor}{gray!12.5}

\newlength{\width}
\newlength{\height}

\def\draftdate{\relax}
\def\mda{\relax}
\def\mua{\relax}
\def\mla{\relax}
\def\draft{
\def\thtystars{******************************}
\def\sixtystars{\thtystars\thtystars}
\typeout{}
\typeout{\sixtystars**}
\typeout{* Draft mode!
         For final version remove \protect\draft\space in source file *}
\typeout{\sixtystars**}
\typeout{}
\def\draftdate{\today}
\def\mua{\marginpar[\boldmath\hfil$\uparrow$]%
                   {\boldmath$\uparrow$\hfil}\color{black}%
                    \typeout{marginpar: $\uparrow$}\ignorespaces}
\def\mda{\color{red}\marginpar[\boldmath\hfil$\downarrow$]%
                   {\boldmath$\downarrow$\hfil}%
                    \typeout{marginpar: $\downarrow$}\ignorespaces}
\def\mla{\marginpar[\boldmath\hfil$\rightarrow$]%
                   {\boldmath$\leftarrow $\hfil}%
                    \typeout{marginpar: $\leftrightarrow$}\ignorespaces}
\def\Mua{\marginpar[\boldmath\hfil$\Uparrow$]%
                   {\boldmath$\Uparrow$\hfil}\color{black}%
                    \typeout{marginpar: $\uparrow$}\ignorespaces}
\def\Mda{\color{red}\marginpar[\boldmath\hfil$\Downarrow$]%
                   {\boldmath$\Downarrow$\hfil}%
                    \typeout{marginpar: $\downarrow$}\ignorespaces}
\def\Mla{\marginpar[\boldmath\hfil\textcolor{red}{$\Rightarrow$}]%
                   {\boldmath\textcolor{red}{$\Leftarrow $}\hfil}%
                    \typeout{marginpar: $\leftrightarrow$}\ignorespaces}
\overfullrule 5pt
\oddsidemargin 15mm
\marginparwidth 29mm
}

\journalname{} % Preprint numbers here
\begin{document}

\title{Precise predictions for same-sign W-boson scattering at the LHC}

\author{Alessandro Ballestrero\thanksref{infnto}
\and
Benedikt Biedermann\thanksref{wurz}
\and
Simon Brass\thanksref{sieg}
\and
Ansgar Denner\thanksref{wurz}
\and
Stefan Dittmaier\thanksref{frei}
\and
Rikkert Frederix\thanksref{tum}
\and
Pietro Govoni\thanksref{mila}
\and
Michele Grossi\thanksref{pavia,ibm}
\and
Barbara J\"ager\thanksref{tub}
\and
Alexander Karlberg\thanksref{uzh}
\and
Ezio Maina\thanksref{infnto,unito}
\and
Mathieu Pellen\thanksref{wurz}
\and
Giovanni Pelliccioli\thanksref{infnto,unito}
\and
Simon Pl\"atzer\thanksref{wien}
\and
Michael Rauch\thanksref{kit}
\and
Daniela Rebuzzi\thanksref{pavia}
\and
J\"urgen Reuter\thanksref{desy}
\and
Vincent Rothe\thanksref{desy}
\and
Christopher Schwan\thanksref{frei}
\and
Hua-Sheng Shao\thanksref{lpthe}
\and
Pascal Stienemeier\thanksref{desy}
\and
Giulia Zanderighi\thanksref{cern}
\and
Marco Zaro\thanksref{nikh}
\and
Dieter Zeppenfeld\thanksref{kit}
}

\institute{INFN, Sezione di Torino, %
Via P. Giuria 1, %
10125 Turin, %
Italy\label{infnto}
\and
Universit\"at W\"urzburg, %
Institut f\"ur Theoretische Physik und Astrophysik,  %
Emil-Hilb-Weg 22, %
97074 W\"urzburg, %
Germany\label{wurz}
\and
Universit\"at Siegen, Department Physik, %
 Walter-Flex-Str.3, %
57068 Siegen, Germany\label{sieg}
\and
Albert-Ludwigs-Universit\"at Freiburg, Physikalisches Institut, %
 Hermann-Herder-Str.\ 3, %
79104 Freiburg, Germany\label{frei}
\and
Technische Universit\"{a}t M\"{u}nchen, %
James-Franck-Str.~1, %
85748 Garching, %
Germany\label{tum}
\and
University and INFN, Milano-Bicocca, %
piazza della Scienza, 3, 20126 Milan, %
Italy\label{mila}
\and
Universit\'a di Pavia, Dipartimento di Fisica and INFN, Sezione di Pavia, %
Via A. Bassi 6, %
27100 Pavia, %
Italy\label{pavia}
\and
IBM Italia s.p.a. %
Circonvallazione Idroscalo, %
20090 Segrate (MI), %
Italy\label{ibm}
\and
University of T\"ubingen,
Institute for Theoretical Physics,
Auf der Morgenstelle 14,
72076 T\"ubingen,
Germany\label{tub}
\and
Universit\"at Z\"urich,
Physik-Institut,
Winterthurerstrasse 190,
8057 Z\"urich,
Switzerland\label{uzh}
\and
Universit\`a di Torino, %
Dipartimento di Fisica, %
Via P. Giuria 1, %
10125 Turin, %
Italy\label{unito}
\and
University of Vienna,
Particle Physics, Faculty of Physics, %
Vienna, %
Austria\label{wien}
\and
Karlsruhe Institute of Technology (KIT), %
Institute for Theoretical Physics, %
76131 Karlsruhe,
Germany\label{kit}
\and
DESY Theory Group, %
Notkestr. 85, %
22607 Hamburg,
Germany\label{desy}
\and
Sorbonne Universit\'e et CNRS, %
Laboratoire de Physique Th\'eorique et Hautes \'Energies (LPTHE), % 
UMR 7589, 
4 place Jussieu, %
75252 Paris Cedex 05, %
France\label{lpthe}
\and
CERN,
Theoretical Physics Department,
1211, Geneva 23,
Switzerland\label{cern}
\and
Nikhef,
Science Park 105,
1098XG Amsterdam,
The Netherlands\label{nikh}
}

\maketitle

\begin{abstract}
  \noindent

  Vector-boson scattering processes are of great importance for the
  current run-II and future runs of the Large Hadron
  Collider. The presence of triple and quartic gauge couplings in the
  process gives access to the gauge sector of the Standard Model (SM) and
  possible new-physics contributions there.  To test any new-physics hypothesis,
  sound knowledge of the SM contributions is necessary,
  with a precision which at least matches the experimental
  uncertainties of existing and forthcoming measurements.  In this
  article we present a detailed study of the vector-boson scattering
  process with two positively-charged leptons and missing transverse
  momentum in the final state. In particular,
  we first carry out a systematic comparison of the various
  approximations that are usually performed for this kind of process
  against the complete calculation, at LO and NLO QCD accuracy. Such a
  study is performed both in the usual fiducial region used by
  experimental collaborations and in a more inclusive phase space,
  where the differences among the various approximations lead to more
  sizeable effects.  Afterwards, we turn to predictions
  matched to parton showers, at LO and NLO: we show that on the one
  hand, the inclusion of NLO QCD corrections leads to more stable
  predictions, but on the other hand the details of the matching and of the
  parton-shower programs cause differences which are considerably
  larger than those observed at fixed order, even in the experimental
  fiducial region.  We conclude with recommendations for
  experimental studies of vector-boson scattering processes.\\[5mm]

  % preprint numbers here
  \noindent
  {\small \underline {Preprint numbers:} 

  \vspace{0.1cm}
  
  DESY 18-025, FR-PHENO-2018-003, KA-TP-05-2018, 
  
  MCnet-18-06, Nikhef/2018-012, UWTHPH-2018-12, 
  
  VBSCan-PUB-01-18, ZU-TH-08/18}
\end{abstract}
\thispagestyle{empty}
\vfill
\newpage
\setcounter{page}{1}

%\tableofcontents
\newpage

\section{Introduction}
Vector-boson scattering (VBS) at a hadron collider 
usually refers to the interaction of massive vector bosons ($\PW^\pm,\PZ$),
radiated by partons (quarks) of the incoming protons, 
which in turn are deflected from the beam direction 
and enter the volume of the particle detectors.
As a consequence, the typical signature of VBS events
is characterised by two energetic jets 
and four fermions,
originating from the decay of the two vector bosons.
Among the possible diagrams,
the scattering process can be mediated by a Higgs boson.
The interaction of longitudinally polarised bosons is of particular interest, 
because the corresponding matrix elements feature unitarity cancellations 
that strongly depend on the actual structure of the Higgs sector of the Standard Model (SM).
A detailed study of this class of processes will therefore further constrain the Higgs couplings 
at a very different energy scale with respect to the Higgs-boson mass,
and hint at, or exclude, non-Standard Model behaviours.

The VBS process involving two same-sign $\PW$ bosons has the largest signal-to-background ratio of all the VBS processes at the Large Hadron Collider (LHC):
evidence for it was found at the centre-of-mass energy of $8~\TeV$~\cite{Aad:2014zda,Khachatryan:2014sta,Aaboud:2016ffv},
and it has been recently measured
at $13\TeV$ as well~\cite{Sirunyan:2017ret}.
Presently, the measurements of VBS processes are limited by statistics, but the situation will change in the near future.
On the theoretical side, 
it is thus of prime importance to provide predictions with systematic uncertainties
at least comparable to the current and envisaged experimental precision~\cite{CMS:2016rcn,ATL-PHYS-PUB-2017-023}.

$\PW^+\PW^+$ scattering is the simplest VBS process to calculate, 
because the double-charge structure of the leptonic final state 
limits the number of partonic processes and total number of Feynman diagrams for each process.
Nonetheless, it possesses all features of VBS at the LHC and is thus representative of other VBS signatures.
Therefore, it is the ideal candidate for a comparative study of the different simulation tools.

In the last few years, several next-to-leading-order (NLO) computations have become available for both the VBS process~\cite{Jager:2006zc,Jager:2006cp,Bozzi:2007ur,Jager:2009xx,Jager:2011ms,Denner:2012dz,Rauch:2016pai} and its QCD-induced irreducible background process~\cite{Rauch:2016pai,Melia:2010bm,Melia:2011gk,Campanario:2013gea,Baglio:2014uba}.
All these VBS computations rely on various approximations, typically neglecting contributions which are expected to be small in realistic experimental setups~\cite{Denner:2012dz,Oleari:2003tc}.
Recently, the complete NLO corrections to $\PW^+\PW^+$ have been evaluated in Ref.~\cite{Biedermann:2017bss}, 
making it possible for the first time to study in detail the quality of the VBS approximations at NLO QCD.\footnote{Preliminary results of the present study have already been made public in Ref.~\cite{Anders:2018gfr}. 
A similar study has also appeared very recently for the electroweak (EW) production of a Higgs boson in association with 3 jets \cite{Campanario:2018ppz}.}

This article starts with the definition of the VBS process in Sec.~\ref{sec:definition}, before describing the approximations of the various computer 
codes in Sec.~\ref{sec:details}. In Sec.~\ref{sec:LO} a leading-order (LO) study 
of the different contributions which lead to the production of two same-sign $\PW$ bosons and 
two jets is performed. In the same section predictions for VBS from different tools are compared at 
the level of the cross section and differential distributions.
The comparison is extended to the
NLO corrections to VBS in Sec.~\ref{sec:NLO}. The effect of the inclusion of matching LO and NLO computations to parton shower (PS) is 
discussed in Sec.~\ref{sec:matching}. Finally,
Sec.~\ref{sec:conclusion} contains a summary of the article and concluding remarks.

\section{Definition of the process}
\label{sec:definition}
\begin{figure*}
\begin{center}
          \includegraphics[width=0.30\linewidth]{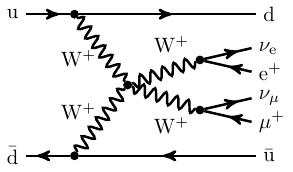}
          \raisebox{.5ex}{\includegraphics[width=0.35\linewidth]{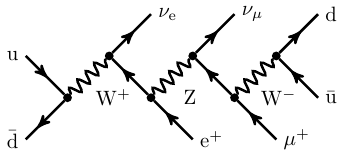}}
          \raisebox{-1.8ex}{\includegraphics[width=0.32\linewidth]{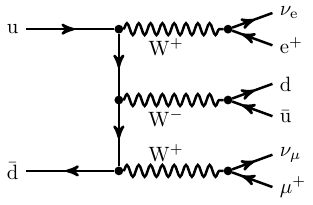}}
\end{center}
        \caption{Sample tree-level diagrams that contribute to the process $\Pp\Pp\to\mu^+\nu_\mu\Pe^+\nu_{\Pe}\Pj\Pj$ at order $\mathcal{O}{\left(\alpha^{6}\right)}$.
        In addition to typical VBS contributions (left), this order also possesses $s$-channel contributions such as decay chain (middle) and tri-boson contributions (right).}
\label{diag:LO}
\end{figure*}

The scattering of two positively-charged $\PW$ bosons with their subsequent decay into different-flavour leptons 
can proceed at the LHC through the partonic process:
\begin{equation}
\Pp\Pp\to\mu^+\nu_\mu{\rm e}^+\nu_{\rm e}\,\Pj\Pj+\mathrm{X}.
\end{equation}

At LO, this process can proceed via three different coupling-order combinations:
$\mathcal{O}{\left(\alpha^{6}\right)}$, $\mathcal{O}{\left(\alphas^{2}\alpha^{4}\right)}$, and $\mathcal{O}{\left(\alphas\alpha^{5}\right)}$.
The first, commonly referred to as EW contribution or VBS,\footnote{The name VBS is used even though not all Feynman diagrams involve the scattering
of vector bosons.} receives contributions from Feynman diagrams such as those depicted in Fig.~\ref{diag:LO}:
in addition to genuine VBS contributions (left diagram), it also features $s$-channel contributions with non-resonant vector bosons 
(centre diagram) or from
triple-boson production (right diagram).
Note that $s$-, $t$-, and $u$-channel contributions are defined according to the quark lines.
The $s$-channel denotes all Feynman diagrams where the two initial-state partons are connected by a continuous fermion line, while for the $t$- and $u$-channel the fermion lines connect initial state quarks to final state quarks.
The $u$-channel refers to contributions with crossed fermion lines with respect to $t$-channel, which appears for identical (anti-)quarks in the final state.
The $s$-channel contributions play a particular role in the study of the various contributions in Sec.~\ref{subsec:contributions}.
When using approximations, care must be taken that only gauge-invariant subsets are considered to obtain physically meaningful results.
We discuss the commonly-used possible choices in detail in the next section.

The second coupling combination of order $\mathcal{O}{\left(\alpha_{\rm s}^2\alpha^{4}\right)}$ corresponds to diagrams with a gluon connecting the two quark lines, and with the $\PW$ bosons 
radiated off the quark lines. Because of the different colour structure, this contribution features a 
different kinematic behaviour than VBS.
Nonetheless it shares the same final state, and therefore constitutes an irreducible background to the EW process.

Finally, the third contribution of order $\mathcal{O}{\left(\alpha_{\rm s}\alpha^{5}\right)}$ is the interference of the two types of amplitudes described above.
It is non-zero only for those partonic sub-processes which involve identical quarks or anti-quarks.
Such a contribution is usually small ($3\%$) within typical experimental cuts~\cite{Biedermann:2017bss}.

In the rest of this article, the notations LO or NLO(-QCD) without any specification of coupling powers refer to the contributions at order $\mathcal{O}{\left(\alpha^{6}\right)}$ and $\mathcal{O}{\left(\alpha_{\rm s}\alpha^{6}\right)}$, respectively.

In experimental measurements, special cuts, called VBS cuts, are designed to enhance the EW contribution over the QCD one and to suppress the interference.
These cuts are based on the different kinematical behaviour of the contributions.
The EW contribution is characterised by two jets with large rapidities as well as a large di-jet invariant mass.
The two $\PW$ bosons are mostly produced centrally.
This is in contrast to the QCD contribution which favours jets in the central region.
Therefore, the event selection usually involves rapidity-difference and invariant-mass cuts for the jets.
Note that, as pointed out in Ref.~\cite{Biedermann:2017bss}, when considering full amplitudes, the separation between EW and QCD production becomes ill
defined.
Hence, combined measurements which are theoretically better defined should be preferably performed by the experimental collaborations at the LHC.

\section{Details of the calculations}
    \label{sec:details}
    \subsection{Theoretical predictions for VBS}
    We now discuss the various approximations which are implemented in computer programs for the EW contribution at order $\mathcal{O}{\left(\alpha^{6}\right)}$.
    Since we are mostly interested in 
    the scattering of two $\PW$ bosons, which includes the quartic gauge-boson vertex, it may appear justified to approximate the 
    full process by considering just those diagrams which contain the $2\rightarrow 2$ scattering process as a sub-part.
    However, this set of contributions is not gauge invariant.
    In order to ensure gauge invariance, an on-shell projection of
    the incoming and outgoing W bosons should be performed.
    While this can be done in the usual way for the time-like
    outgoing W bosons, the treatment of the space-like W bosons
    emitted from the incoming quarks requires some care.
    Following Refs.~\cite{Kuss:1995yv,Accomando:2006hq} these W-boson lines can be split,
    the W bosons entering the scattering process can be projected
    on-shell, and the emission of the W bosons from the quarks can be
    described by vector-boson luminosities.
    Such an approximation is usually called effective vector-boson
    approximation (EVBA) \cite{Dawson:1984gx,Duncan:1985vj,Cahn:1983ip}.

    An improvement of such an approximation consists 
    in considering all $t$- and $u$-channel diagrams and squaring them separately, neglecting interference contributions between the two classes.
    These interferences are expected to be small in the VBS fiducial region, as they are both phase-space and colour suppressed~\cite{Oleari:2003tc,Denner:2012dz}.
    The $s$-channel squared diagrams and any interferences between them and the  $t/u$-channels are also discarded.
    This approximation is often called $t$-/$u$- approximation, VBF, or even VBS approximation.
    We adopt the latter denomination in the following.
    This approximation is gauge-invariant, a fact that can be appreciated by considering the two incoming quarks as belonging to two different copies of the $\rm{SU}(3)$ gauge group.

    A further refinement is to add the squared matrix element of the $s$-channel contributions to the VBS approximation.

    The approximations performed at LO can be extended when NLO QCD corrections to the order $\mathcal{O}{\left(\alpha^{6}\right)}$ are computed.
    The VBS approximation can be extended at NLO in a straightforward manner for what concerns the virtual contributions.
    For the real-emission contributions special care must be taken for the gluon-initiated processes.
    The initial-state gluon and initial-state quark 
        must not couple together, otherwise infrared (IR) divergences proportional to $s$-channels appear, 
         which do not match with the ones found in the virtual contributions.
         The subset of diagrams where all couplings of the initial state gluon to initial state quark are neglected forms a gauge-invariant subset, with the same argument presented above.

    A further refinement is to consider the full real contributions, which include all interferences, and part of the virtual.
    In particular one can consider only one-loop amplitudes where there is no gluon exchange between the two quark lines and 
    assuming a cancellation of the IR poles.

    When considering the full NLO corrections of order $\mathcal{O}{\left(\alpha_{\rm s}\alpha^{6}\right)}$, besides real and virtual QCD corrections
    to the EW tree-level contribution of order
    $\mathcal{O}{\left(\alpha^{6}\right)}$
    also real and virtual EW corrections to the LO interference
    of order $\mathcal{O}{\left(\alpha_{\rm s}\alpha^{5}\right)}$
    have to be taken into account. Since some loop diagrams contribute
    to both types of corrections, QCD and EW corrections cannot be
    separated any more on the basis of Feynman diagrams, and the
    cancellation of IR singularities requires the inclusion of all of them \cite{Biedermann:2017bss}.
    \subsection{Description of the programs used}
        \label{subsec:codedescr}
    In the following, the codes employed throughout this article and the approximations implemented in each of them are discussed:

    \begin{itemize}
      \item {\sc Phantom}~\cite{Ballestrero:2007xq} is a dedicated tree-level Monte Carlo for six-parton final states 
      at $\Pp \Pp,\, \Pp\bar{\Pp}$, and $\Pe^+\Pe^-$ colliders at orders $\mathcal O(\alpha^6)$ and $\mathcal O(\alphas^2\alpha^4)$ including interferences between the two sets of diagrams.
    It employs complete tree-level matrix elements in the complex-mass scheme~\cite{Denner:1999gp,Denner:2005fg,Denner:2006ic} computed via the modular helicity formalism~\cite{Ballestrero:1999md,Ballestrero:1994jn}.
    The integration uses a multi-channel approach~\cite{Berends:1984gf} and an adaptive strategy~\cite{Lepage:1977sw}.
    {\sc Phantom} generates unweighted events at parton level for both the SM and a few instances of beyond the Standard Model (BSM) theories.

      \item WHIZARD~\cite{Kilian:2007gr,Moretti:2001zz} is a multi-purpose event generator with LO matrix-element generator O'Mega. 
      For QCD amplitudes it uses the colour flow formalism~\cite{Kilian:2012pz}.
      For NLO QCD calculations~\cite{Nejad:2016bci}, where WHIZARD is in the final validation phase, it provides FKS subtraction terms~\cite{Frixione:1995ms,Frixione:1997np}, while virtual matrix elements are provided externally by OpenLoops~\cite{Cascioli:2011va} or Recola~\cite{Actis:2012qn,Actis:2016mpe}.
      Furthermore, WHIZARD can automatically provide POWHEG matching to parton shower~\cite{Reuter:2016qbi}.
      WHIZARD allows to simulate a huge number of BSM models as well, in particular in terms of higher-dimensional operators for VBS processes including means to provide unitarity limits~\cite{Alboteanu:2008my,Kilian:2014zja}.
      
    \item The program {\sc Bonsay}~\cite{Dittmaier:2018zzz} consists of a general-purpose Monte Carlo integrator and matrix elements taken from different sources:
    Born matrix elements are adapted from the program {\sc Lusifer}~\cite{Dittmaier:2002ap}, which have been generalised to calculate also real matrix elements.
    Virtual matrix elements have been calculated using an in-house matrix-element generator.
    One-loop integrals are evaluated using the {\sc Collier} library~\cite{Denner:2014gla,Denner:2016kdg}.
    For the results presented here, it uses the VBS approximation at LO and NLO.
    The virtual corrections are additionally approximated using a double-pole approximation where the final state leptons are assumed to originate from the decay of two resonant W-bosons; see Ref.~\cite{Denner:2000bj} for a
detailed description and Ref.~\cite{Dittmaier:2015bfe} for the on-shell projection used.
    At LO the exact matrix elements can also be used.

      \item The {\sc Powheg-Box}~\cite{Nason:2004rx,Frixione:2007vw,Alioli:2010xd} is a framework for matching NLO-QCD calculations with parton showers.
    It relies on the user providing the matrix elements and Born phase space, but automatically constructs FKS \cite{Frixione:1995ms} subtraction terms and the phase space corresponding to the real emission.
    For the VBS processes all matrix elements are being provided by a previous version of {\sc VBFNLO}~\cite{Arnold:2008rz, Arnold:2011wj, Baglio:2014uba} and hence the approximations used in the {\sc Powheg-Box} are similar to those used in {\sc VBFNLO}.

      \item {\sc VBFNLO}~\cite{Arnold:2008rz, Arnold:2011wj, Baglio:2014uba} is a flexible
        parton-level Monte Carlo for processes with EW bosons. It
        allows the calculation of VBS processes at NLO QCD in the VBS
        approximation, with process IDs between 200 and 290. Same-sign
        $\PW^+\PW^+jj$ production is provided via the process ID 250. The corresponding
        $s$-channel contributions are available separately as tri-boson processes with
        semi-leptonic decays, with process IDs between 401 and 492. For the
        final state studied in this article, only $\PW^+\PW^+\PW^-$
        production with a hadronically decaying $\PW^-$ boson, process ID 432,
        can contribute. These can simply be added on top of the VBS
        contribution. Interferences between the two are therefore neglected.
        The usage of leptonic tensors in the calculation, pioneered in
        Ref.~\cite{Jager:2006zc}, thereby leads to a significant speed improvement over
        automatically generated code.  Besides the SM, also a variety of
        new-physics models including anomalous couplings of the Higgs and gauge
        bosons can be simulated.
        
      \item {\sc MadGraph5\_aMC@NLO}~\cite{Alwall:2014hca} (henceforth {\sc MG5\-\_aMC}) is an automatic meta-code (a code that generates codes) which makes it possible to simulate any scattering process
          including NLO QCD corrections both at fixed order and including matching to parton showers, using the {\sc MC@NLO}\ method~\cite{Frixione:2002ik}. It makes use of the subtraction method by Frixione, Kunszt and Signer (FKS)~\cite{Frixione:1995ms,
            Frixione:1997np} (automated in the module {\sc MadFKS}~\cite{Frederix:2009yq,
            Frederix:2016rdc}) for regulating IR singularities. The computations of one-loop amplitudes are carried out by switching dynamically between
            two integral-reduction techniques, OPP~\cite{Ossola:2006us} or Laurent-series expansion~\cite{Mastrolia:2012bu},
            and tensor-integral reduction~\cite{Passarino:1978jh,Davydychev:1991va,Denner:2005nn}. These have been automated in the module {\sc MadLoop}~\cite{Hirschi:2011pa}, which
            in turn exploits {\sc CutTools}~\cite{Ossola:2007ax}, {\sc Ninja}~\cite{Peraro:2014cba,
            Hirschi:2016mdz}, {\sc IREGI}~\cite{ShaoIREGI}, or {\sc Collier}~\cite{Denner:2016kdg}, together with an in-house 
            implementation of the {\sc OpenLoops} optimisation~\cite{Cascioli:2011va}. Finally, scale and PDF uncertainties can be obtained in an exact manner via reweighting
            at negligible additional CPU cost~\cite{Frederix:2011ss}.\\
            The simulation of VBS at NLO-QCD accuracy can be performed by issuing the following commands in the program interface:
    \begin{verbatim}
set complex_mass_scheme
import model loop_qcd_qed_sm_Gmu
generate p p > e+ ve mu+ vm j j QCD=0 [QCD]
output
    \end{verbatim}
      With these commands the complex-mass scheme is turned on, then the NLO-capable model is loaded,\footnote{Despite
                the {\tt loop\_qcd\_qed\_sm\_Gmu} model also includes NLO counterterms for computing EW corrections, it is not yet possible to compute such corrections
            with the current public version of the code.} finally the process code is generated (note the {\tt QCD=0} syntax to select the purely-EW process)
            and written to disk. No approximation is performed for the Born and real-emission matrix elements. 
            Only strongly-in\-ter\-act\-ing particles circulating in the loops are generated for the virtual matrix element.
            The version capable of computing both QCD and EW corrections will be released in the future.
            Such an approximation is equivalent to the assumption that the finite part of
            those loops which feature EW bosons is zero. In practice, since a part of the contribution to the single pole is also missing, the internal 
            pole-cancellation check at run time has to be turned off, by setting the value of the {\tt IR\-Pole\-Check\-Threshold} and 
            {\tt Precision\-Virtual\-At\-Run\-Time} parameters in the {\tt Cards\-/FKS\_\-params.dat} file to -1.

    \item The program {\sc MoCaNLO+Recola} is made of a flexible Monte Carlo program dubbed {\sc MoCaNLO} and of the matrix-element generator {\sc Recola}~\cite{Actis:2012qn,Actis:2016mpe}.
    It can compute arbitrary processes at the LHC with both NLO QCD and EW accuracy in the SM.
    This is made possible by the fact that {\sc Recola} can compute arbitrary processes at tree and one-loop level in the SM.
    To that end, it relies on the {\sc Collier} library \cite{Denner:2014gla,Denner:2016kdg} to numerically evaluate the one-loop scalar and tensor integrals.
    In addition, the subtraction of the IR divergences appearing in the real corrections has been automatised thanks to the Catani--Seymour dipole formalism for both QCD and QED \cite{Catani:1996vz,Dittmaier:1999mb}.
    The code has demonstrated its ability to compute NLO corrections for high-multiplicity processes up to $2 \to 7$ \cite{Denner:2015yca,Denner:2016wet}.
    In particular the full NLO corrections to VBS and its irreducible background \cite{Biedermann:2016yds,Biedermann:2017bss} have been obtained thanks to this tool.
    One key aspect for these high-multiplicity processes is the fast integration which is ensured by using similar phase-space mappings to those of Refs.~\cite{Berends:1994pv,Denner:1999gp,Dittmaier:2002ap}. 
    In {\sc MoCaNLO+Recola} no approximation is performed neither at LO nor at NLO.
    It implies that, also contributions stemming from EW corrections to the interference are computed.
            
    \end{itemize}

    We conclude this section by summarising the characteristics of the various codes in Tab.~\ref{tab:wg1_codes}.
    In particular, it is specified whether
    \begin{itemize}
        \item all $s$-,$t$-,$u$-channel diagrams are included;
        \item interferences between diagrams of different types are included at LO;
        \item diagrams which do not feature two resonant W bosons are included;
        \item the so-called non-factoris\-able (NF) QCD corrections, \emph{i.e.}\ the corrections where (real or virtual) gluons are exchanged between different quark lines,
            are included;
        \item EW corrections to the interference of order $\mathcal O (\alpha^5\alphas)$ are included.
        These corrections are of the same order as the NLO QCD corrections to the contribution of order $\mathcal O (\alpha^6$) term.
    \end{itemize}

    \begin{table*}
        \footnotesize
        \begin{tabularx}{\textwidth}{c|c|X|X|X|X|X}
            Code  &  $\mathcal O(\alpha^6)$ $s, t, u$  &  $\mathcal O(\alpha^6)$ interf.  &  Non-res.  & NLO &  NF QCD  &  EW corr. to order $\mathcal O(\alphas \alpha^5)$  \\
            \hline
            \hline
            {\sc Phantom}       &  $s,t,u$  &  Yes      &  Yes              &  No    & -        & - \\
            {\sc WHIZARD}       &  $s,t,u$  &  Yes      &  Yes              &  No    & -        & - \\
            {\sc Bonsay}        &  $t,u$    &  No       &  Yes, virt. No    &  Yes   & No       &  No  \\
            {\sc Powheg}        &  $t,u$    &  No       &  Yes              &  Yes   & No       &  No  \\
            {\sc VBFNLO}        &  $s,t,u$  &  No       &  Yes              &  Yes   & No       &  No  \\
            {\sc MG5\_aMC}      &  $s,t,u$  &  Yes      &  Yes              &  Yes   & virt. No &  No \\
            {\sc MoCaNLO+Recola}&  $s,t,u$  &  Yes      &  Yes              &  Yes   & Yes      &  Yes  \\
        \end{tabularx}
        \caption{\label{tab:wg1_codes} Summary of the different properties of the computer programs employed in the comparison.}
    \end{table*}
    \subsection{Input parameters}
        \label{subsec:inputpar}
    The hadronic scattering processes are simulated at the LHC with a centre-of-mass energy $\sqrt s = 13 \TeV$.
    The NNPDF~3.0 parton distribution functions~(PDFs)~\cite{Ball:2014uwa} with five massless flavours,\footnote{For the process considered, no bottom (anti-)quarks appear in the initial or final state at LO and NLO, as they would lead to top quarks, and not light jets, in the final state.} 
    NLO-QCD evolution, and a strong coupling constant $\alphas\left( \MZ \right) = 0.118$\footnote{Note that the {\sc Powheg-Box} uses its own implementation of the two loop running for $\alpha_{\rm s}$.} are employed.\footnote{The corresponding identifier {\tt lhaid} in the program LHAPDF6~\cite{Buckley:2014ana} is 260000.}
    Initial-state collinear singularities are factorised according to the ${\overline{\rm MS}}$ scheme, consistently with what is done in NNPDF.

    For the massive particles, the following masses and decay widths are used:
    \begin{alignat}{2}
                      \Mt   &=  173.21\GeV,       & \quad \quad \quad \Gt &= 0 \GeV,  \nonumber \\
                    \MZOS &=  91.1876\GeV,      & \quad \quad \quad \GZOS &= 2.4952\GeV,  \nonumber \\
                    \MWOS &=  80.385\GeV,       & \GWOS &= 2.085\GeV,  \nonumber \\
                    M_{\rm H} &=  125.0\GeV,       &  \GH   &=  4.07 \times 10^{-3}\GeV.
    \end{alignat}
    The measured on-shell (OS) values for the masses and widths of the W and Z bosons are converted into pole values for the gauge bosons ($V=\PW,\PZ$) according to Ref.~\cite{Bardin:1988xt},
    \begin{equation}
    \begin{split}
            M_V &= \MVOS/\sqrt{1+(\GVOS/\MVOS)^2}\,, \\
       \Gamma_V &= \GVOS/\sqrt{1+(\GVOS/\MVOS)^2}.
    \end{split}
    \end{equation}
    The EW coupling is renormalised in the $G_\mu$ scheme \cite{Denner:2000bj} according to 
    \begin{equation}
    \alpha =  \frac{\sqrt{2}}{\pi} G_{\mu} M_{\rm W}^2 \left(1-\frac{M_{\rm W}^2}{M_{\rm Z}^2} \right),
    \end{equation}
    with
    \begin{equation}
        G_{\mu}    = 1.16637\times 10^{-5}\GeV^{-2},
    \end{equation}
    and where $M_V^2$ corresponds to the real part of the squared pole mass.
    The numerical value of $\alpha$, corresponding to the choice of input parameters is
    \begin{equation}
     1/\alpha = 132.3572\ldots\,.
    \end{equation}
    The Cabibbo--Kobayashi--Maskawa matrix is assumed to be diagonal, meaning that the mixing between different quark generations is neglected.
    The complex-mass scheme~\cite{Denner:1999gp,Denner:2005fg,Denner:2006ic} is used throughout to treat unstable intermediate particles in a gauge-invariant manner.

    The central value of the renormalisation and factorisation scales is set to 
    \begin{equation}
    \label{eq:defscale}
     \mu_{\rm ren} = \mu_{\rm fac} = \sqrt{p_{\rm T, j_1}\, p_{\rm T, j_2}}, 
    \end{equation}
    defined via the transverse momenta of the two hardest jets (identified with the procedure outlined in the following), 
    event by event.\footnote{By default, the renormalisation and factorisation scales employed in the {\sc Powheg-Box} slightly differ from the 
        ones defined in Eq.~\eqref{eq:defscale}, as the momenta of two final-state quarks in the underlying Born event are
        employed instead of those of the two hardest jets.\label{foot:powheg}} 
    This choice of scale has been shown to provide stable NLO-QCD predictions \cite{Denner:2012dz}.

    Following experimental measurements \cite{Aad:2014zda,Aaboud:2016ffv,Khachatryan:2014sta,CMS:2017adb}, the event selection used in the present study is:

    \begin{itemize}
        \item The two same-sign charged leptons are required to fulfil cuts on transverse momentum, rapidity, and separation in the rapidity--azimuthal-angle separation, 
            \begin{align}
            \label{cut:1}
             \ptsub{\Pl} >  20\GeV,\qquad |y_{\Pl}| < 2.5, \qquad \Delta R_{\Pl\Pl}> 0.3\,.
            \end{align}
        \item The total missing transverse momentum, computed from the vectorial sum of the transverse momenta of the two neutrinos, is required to be
            \begin{align}
            \label{cut:2}
              p_{\rm T, miss} >  40\GeV\,.
            \end{align}
        \item QCD partons (light quarks and gluons) are clustered together using the anti-$k_T$ algorithm~\cite{Cacciari:2008gp}, possibly using the {\sc FastJet} implementation~\cite{Cacciari:2011ma}, with distance parameter $R=0.4$.
        We impose cuts on the jets' transverse momenta, rapidities, and their separation from leptons,  
            \begin{align}
            \label{cut:3}
             \ptsub{\Pj} >  30\GeV, \qquad |y_\Pj| < 4.5, \qquad \Delta R_{\Pj\Pl} > 0.3 \,.
            \end{align}
            VBS cuts are applied on the two jets with largest transverse momentum, unless otherwise stated. In particular, we impose a cut on the 
             in\-vari\-ant mass of the di-jet system, as well as on the rapidity separation of the two jets,          
            \begin{align}
            \label{cut:4}
             m_{\Pj \Pj} >  500\GeV,\qquad |\Delta y_{\Pj \Pj}| > 2.5, 
            \end{align}
            if not explicitly stated otherwise. 
        \item When EW corrections are computed, real photons and charged fermions are clustered together using the anti-$k_T$ algorithm with
            radius parameter $R=0.1$. In this case, leptons and quarks are understood as {\it dressed fermions}.
    \end{itemize}

\section{Leading-order study}
    \label{sec:LO}
    \subsection{Contributions}
        \label{subsec:contributions}
    In the present section, the cross sections and distributions are obtained without applying the VBS cuts on the variables $m_{\Pj\Pj}$ and $|\Delta y_{\Pj\Pj}|$, 
    Eq.~(\ref{cut:4}).
    In Tab.~\ref{tab:LOscanXsec}, the cross sections of the three LO contributions are reported.
    The EW, QCD, and interference contributions amount to $57\%$, $37\%$, and $6\%$ of the total inclusive cross section, respectively.
    The QCD contribution does not possess external gluons due to charge conservation.
    Thus the diagrams of order $\mathcal{O}(g_{\rm s}^2g^4)$ only involve gluon exchange between the quark lines.
    This results in a small contribution even if the VBS cuts have not been imposed.
    The interference between EW and QCD contributions is small, due to colour suppression, but not negligible.

    \begin{table}
        \centering
        \begin{tabular}{c|c|c|c}
            Order & $\mathcal{O}(\alpha^6)$ & $\mathcal{O}(\alphas^2\alpha^4)$ & $\mathcal{O}(\alphas\alpha^5)$ \\
            \hline
            \hline
            $\sigma[\rm{fb}]$ & $ 2.292 \pm 0.002 $ & $ 1.477 \pm 0.001 $ & $ 0.223 \pm 0.003 $ \\
    %         {\sc Xxx}&  $ \pm $ & $ \pm $ & $ \pm $
        \end{tabular}
        \caption{\label{tab:LOscanXsec} Cross sections at LO accuracy for the three contributions to the process ${\rm p}{\rm p}\to\mu^+\nu_\mu{\rm e}^+\nu_{\rm e}{\rm j}{\rm j}$, obtained with exact matrix elements.
        These results are for the set-up described in Sec.~\protect\ref{subsec:inputpar} but no cuts on $m_{\Pj\Pj}$ and $|\Delta y_{\Pj\Pj}|$ are applied.
        The uncertainties shown refer to the estimated statistical errors of the Monte Carlo integrations.}
    \end{table}

\begin{figure*}
    \centering
    \includegraphics[scale=0.395]{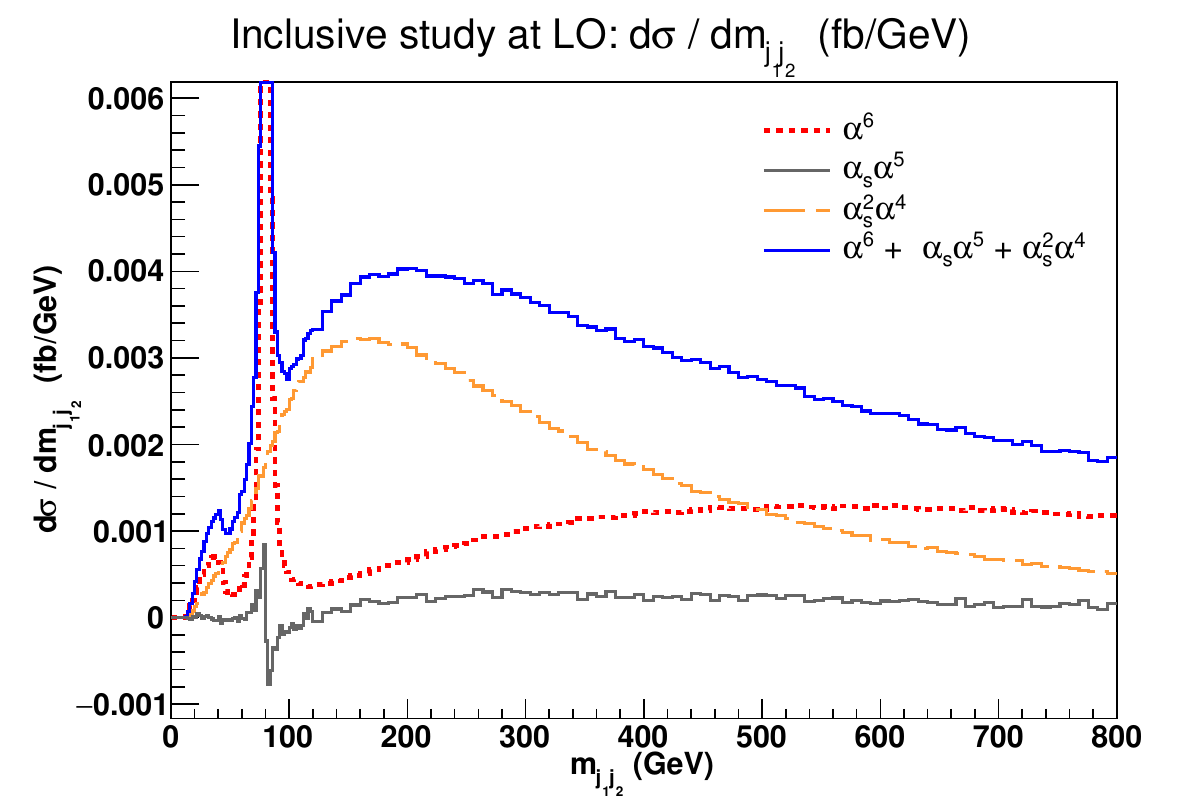}
    \includegraphics[scale=0.395]{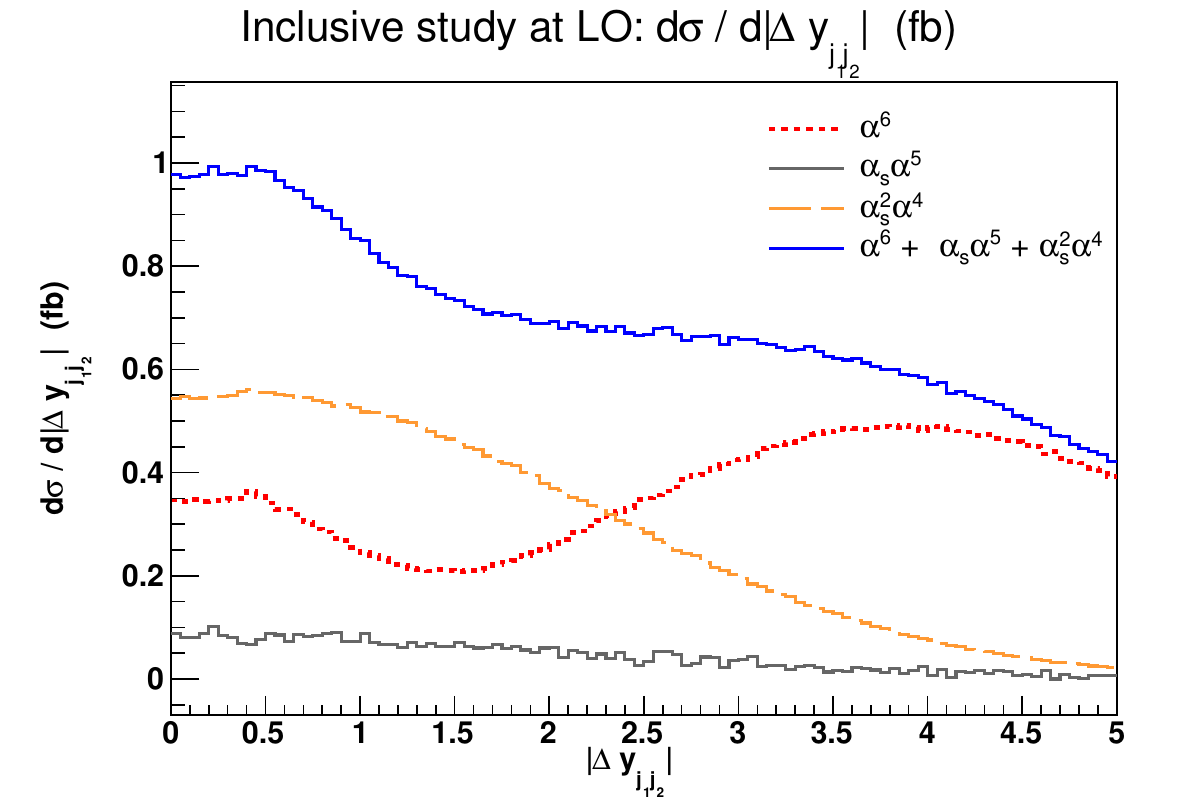}
    \caption{Differential distribution in the di-jet invariant mass $m_{\Pj\Pj}$ (left) and the difference of the jet rapidities $|\Delta y_{\Pj\Pj}|$ (right) for the three LO contributions to the process ${\rm p}{\rm p}\to\mu^+\nu_\mu{\rm e}^+\nu_{\rm e}{\rm j}{\rm j}$.
    The EW contribution is in red, the QCD one in orange, and the interference one in grey.
    The sum of all the contributions is in blue.
    The cuts applied are the ones of Sec.~\protect\ref{subsec:inputpar} but no cuts on $m_{\Pj\Pj}$ and $|\Delta y_{\Pj\Pj}|$ are applied.}
    \label{fig:mjjdyjj_1d}
    \end{figure*}

    In Fig.~\ref{fig:mjjdyjj_1d}, these three contributions are shown separately and summed in the differential distributions in the di-jet invariant mass $m_{\Pj\Pj}$ and the rapidity difference $|\Delta y_{\Pj\Pj}|$.
    For the di-jet invariant-mass distribution (left), one can observe that the EW contribution peaks around an invariant mass of about $80\GeV$.
    This is due to diagrams where the two jets originate from the decay of a W boson (see middle and right diagrams in Fig.~\ref{diag:LO}).
    Note that these contributions are not present in calculations relying on the VBS approximation as these are only $s$-channel contributions.
    The EW contribution becomes dominant for di-jet invariant mass larger than $500\GeV$.
    The same holds true for jet-rapidity difference larger than $2.5$ (right).
    This justifies why cuts on these two observables are used in order to enhance the EW contribution over the QCD one.
    In particular, in order to have a large EW contribution, rather exclusive cuts are required.

\begin{figure*}
    \centering
    \includegraphics[scale=0.395]{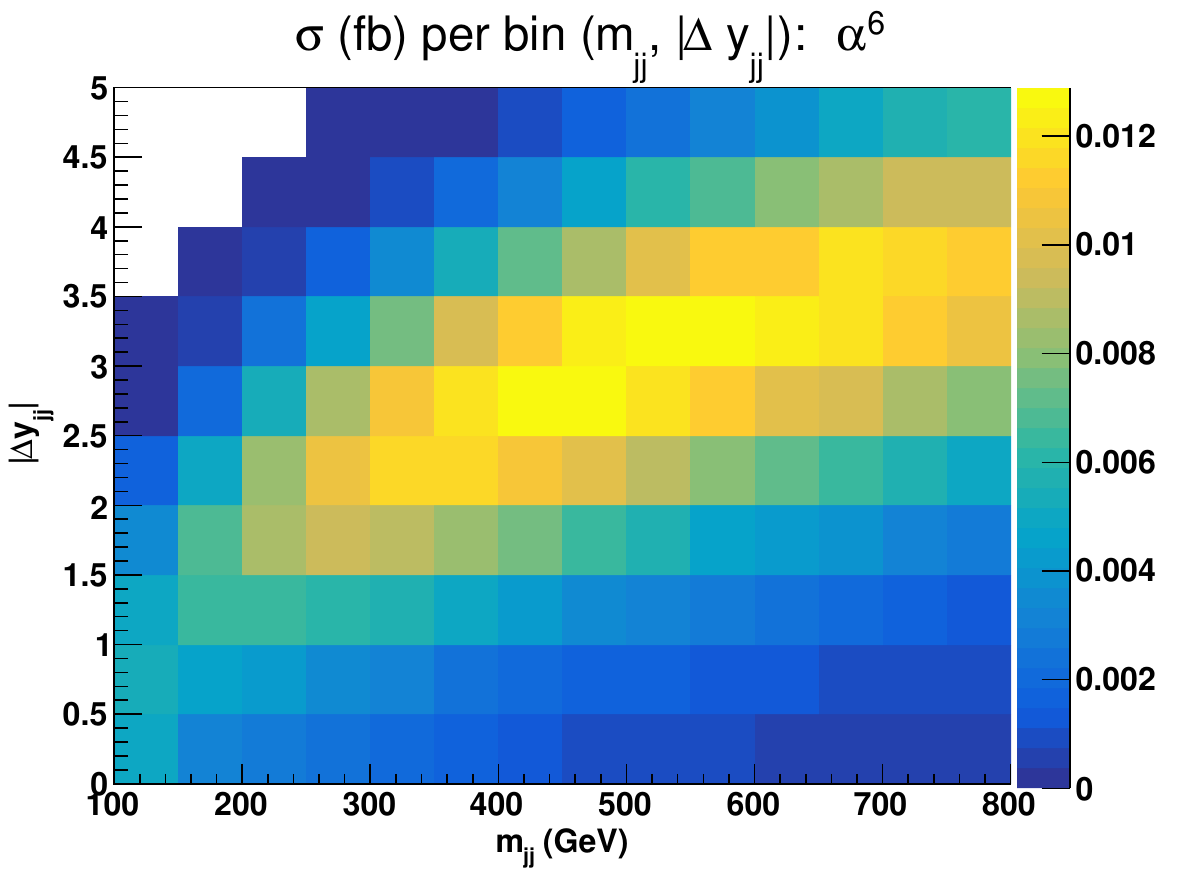}
    \includegraphics[scale=0.395]{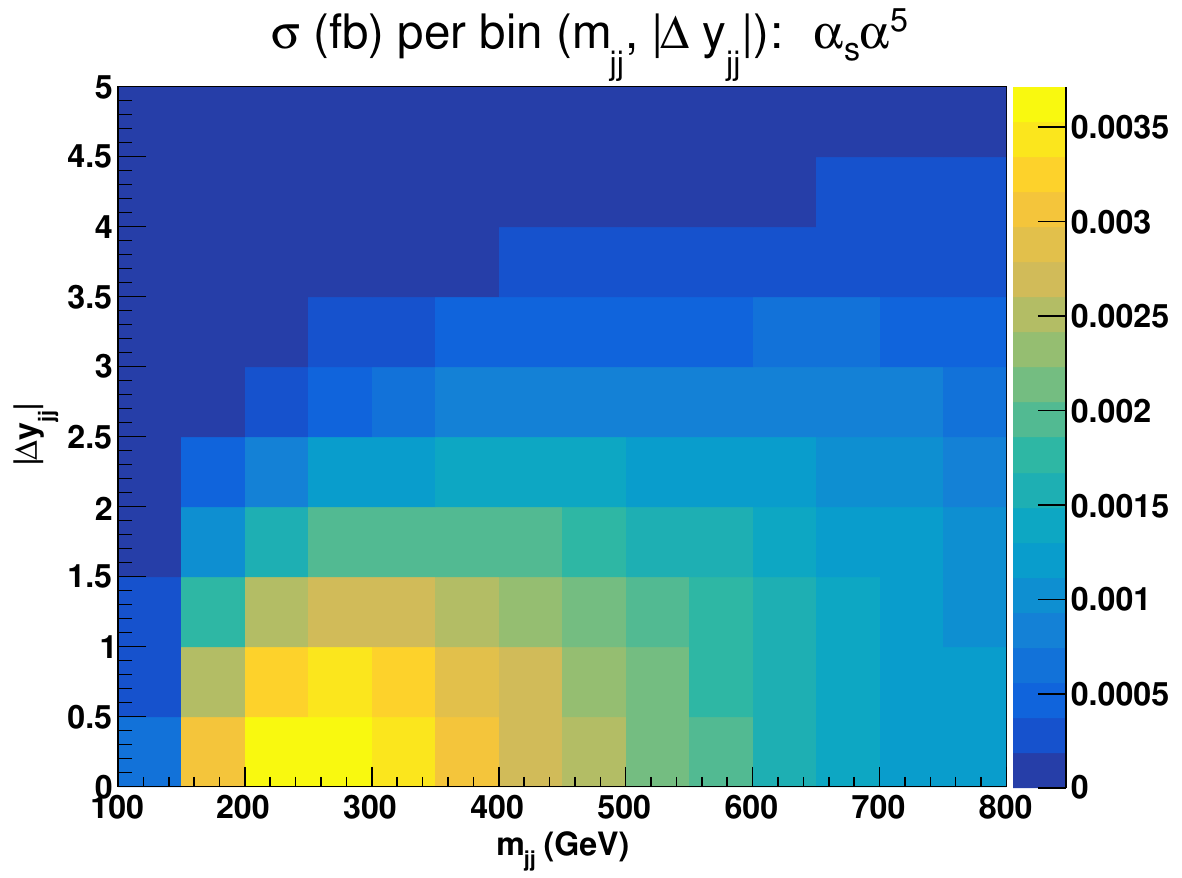}
    \includegraphics[scale=0.395]{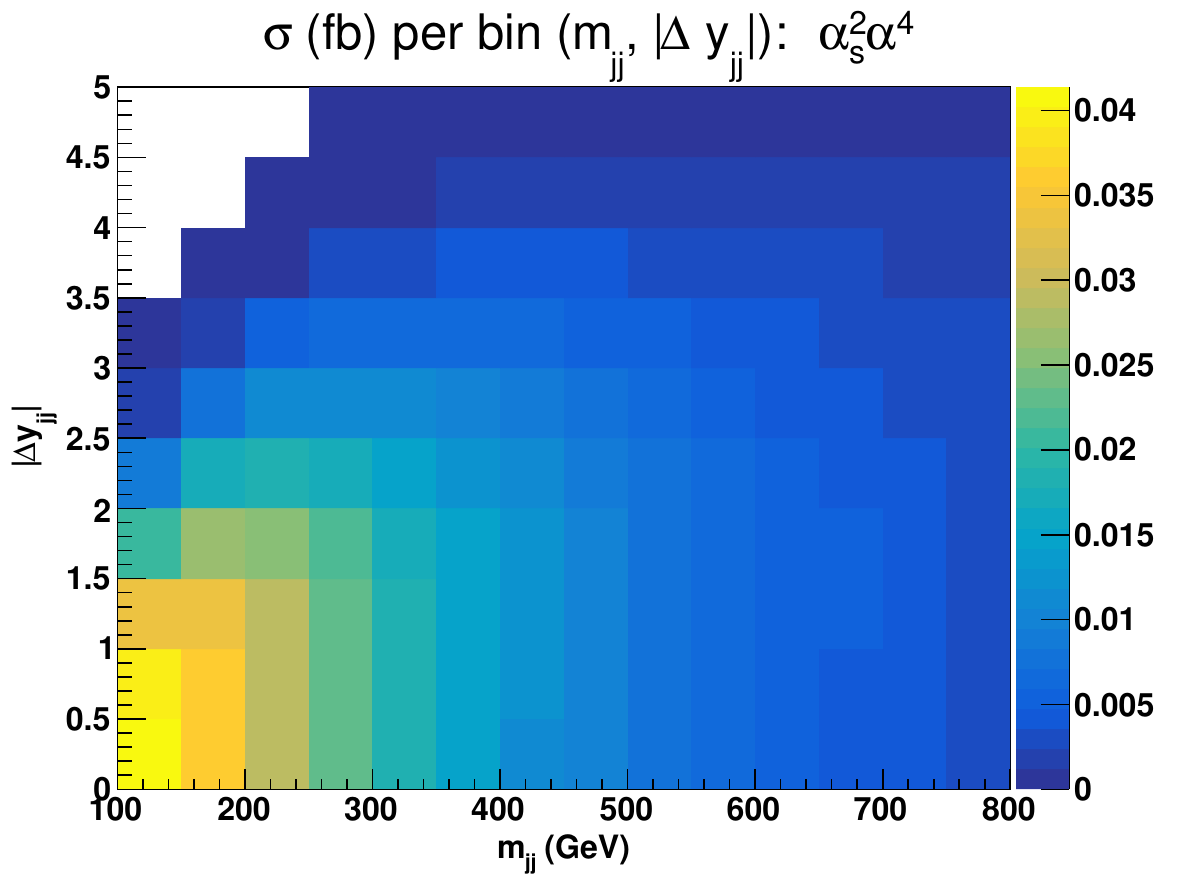}
    \caption{Double-differential distributions in the variables $m_{\Pj\Pj}$ and $|\Delta y_{\Pj\Pj}|$ for the three LO contributions of orders $\mathcal{O}(\alpha^6)$ (top left), $\mathcal{O}(\alphas\alpha^5)$ (top right), and $\mathcal{O}(\alphas^2 \alpha^4)$ (bottom).
    The cuts applied are the ones of Sec.~\protect\ref{subsec:inputpar} but no cuts on $m_{\Pj\Pj}$ and $|\Delta y_{\Pj\Pj}|$ are applied.
    }
    \label{fig:mjjdyjj_2d_LO}
    \end{figure*}

    This can also be seen in Fig.~\ref{fig:mjjdyjj_2d_LO} where the three contributions are displayed as double-differential distributions in the di-jet invariant mass and jet rapidity difference.
    Again, it is clear that the region with low di-jet invariant mass should be avoided in VBS studies as it is dominated by tri-boson contributions.
    This motivates in particular the choice of the cut $m_{\Pj\Pj} > 200\GeV$ for our LO inclusive study in Sec.~\ref{subsec:LOinclusive}.

    \subsection{Inclusive comparison}
        \label{subsec:LOinclusive}
\begin{figure*}
    \centering
    \includegraphics[scale=0.395]{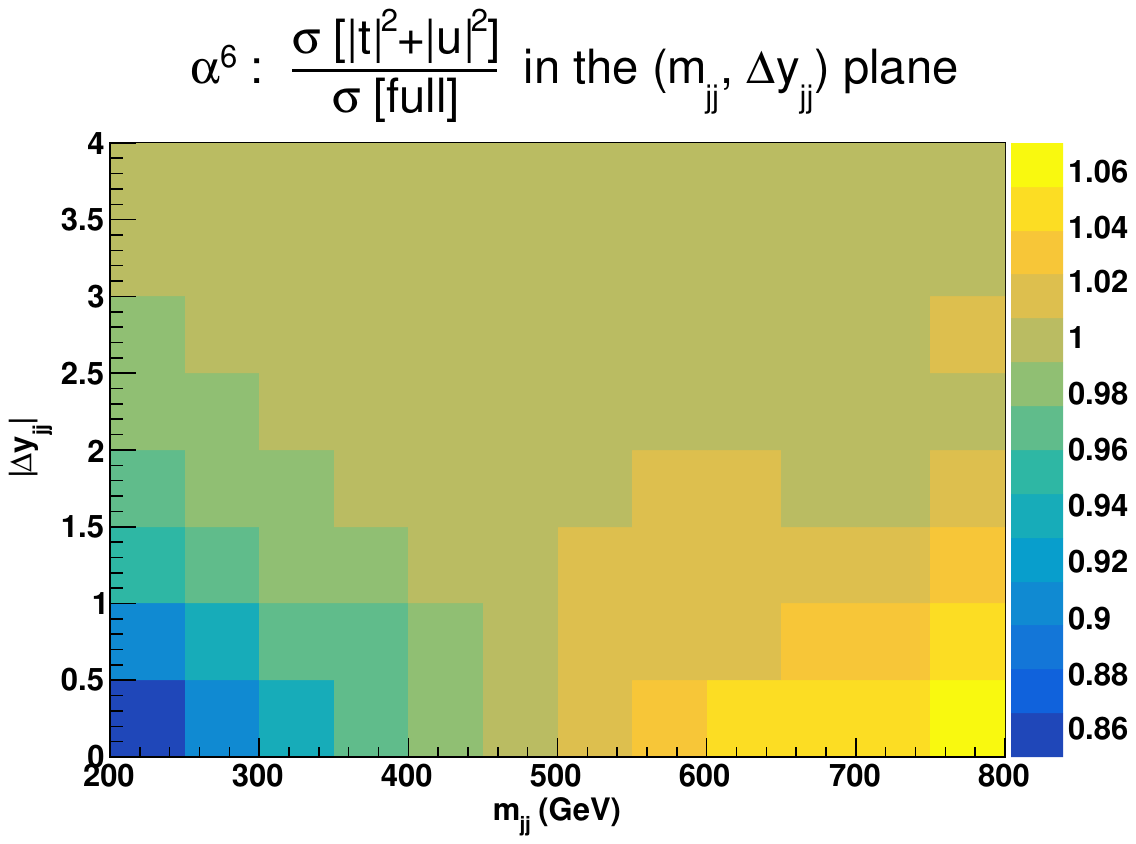}
    \includegraphics[scale=0.395]{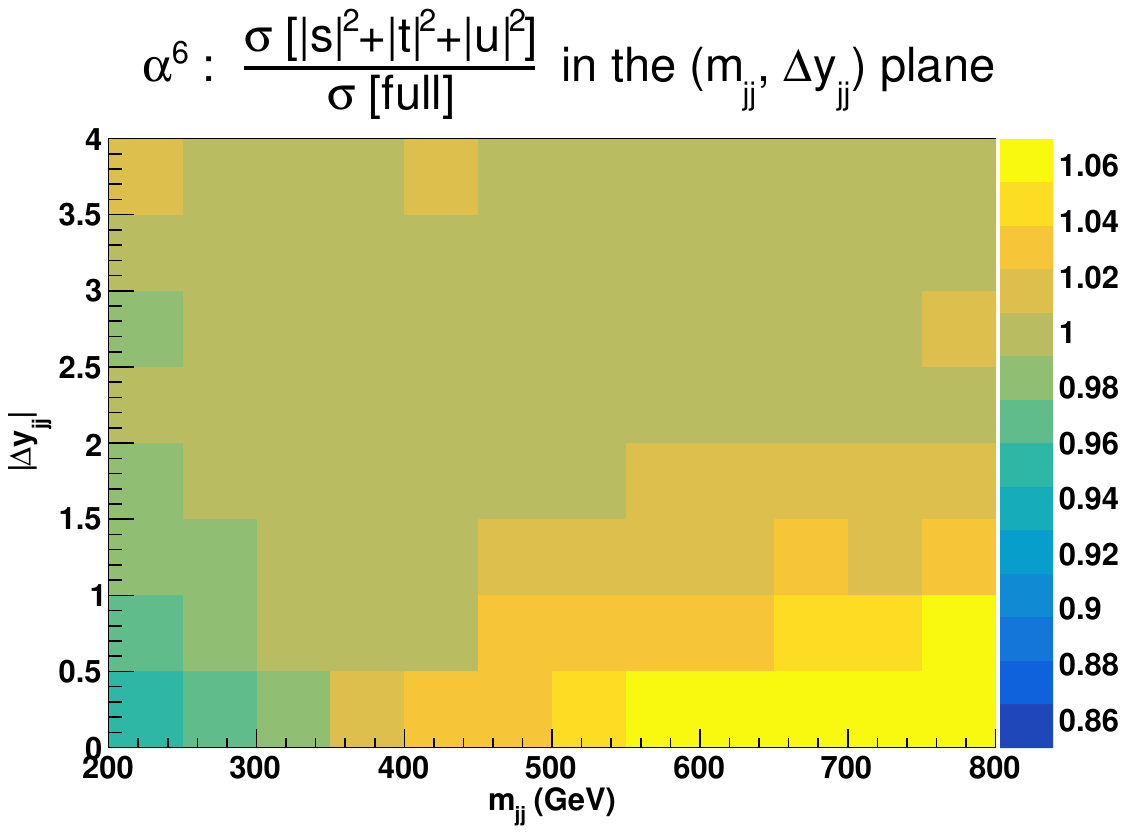}
    \caption{Ratios for double-differential distributions in the variables $m_{\Pj\Pj}$ and $|\Delta y_{\Pj\Pj}|$ at LO \emph{i.e.}\ order $\mathcal{O}(\alpha^6)$ of approximated squared amplitudes over the full matrix element.
    The approximated squared amplitudes are computed as $|\mathcal{A}|^2 \sim |t|^2 + |u|^2$ (left) and $|\mathcal{A}|^2 \sim |s|^2 + |t|^2 + |u|^2$ (right).
    The cuts applied are the ones of Sec.~\protect\ref{subsec:inputpar} and no cuts on $m_{\Pj\Pj}$ and $|\Delta y_{\Pj\Pj}|$ are applied.} 
    \label{fig:ratio2d_LO}
    \end{figure*}

    In Fig.~\ref{fig:ratio2d_LO}, ratios for double-differential cross sections in the variables  $m_{\Pj\Pj}$ and $|\Delta y_{\Pj\Pj}|$ are shown.\footnote{In Fig.~\ref{fig:ratio2d_LO}, the level of the accuracy of the predictions in each bin is around a per mille.}
    Two plots are displayed: the ratios of the $|t|^2 + |u|^2$ and $|s|^2 + |t|^2 + |u|^2$ approximations over the full calculation.
    In the first case, the approximation is good within $\pm10\%$ over the whole range apart from the low invariant-mass region at both low and large rapidity difference.
    The low rapidity-difference region possesses remnants of the tri-bosons contribution that have a di-jet invariant mass around the $\PW$-boson mass.
    It is therefore expected that the $|t|^2 + |u|^2$ approximation fails in this region.
    The second plot, where the $|s|^2 + |t|^2 + |u|^2$ approximation is considered, displays a better behaviour in the previously mentioned region.
    The full calculation is approximated at the level of $\pm5\%$ everywhere apart from the region where $|\Delta y_{\Pj\Pj}| < 2$.

    \subsection{Comparison in the fiducial region}
    In Tab.~\ref{tab:wg1_LOrates}, we report the total rates at LO accuracy at order $\mathcal O (\alpha^6)$ obtained in the fiducial region 
    described in Sec.~\ref{subsec:inputpar}. Two things should be highlighted here: first, despite the different underlying approximations, 
    the two most-distant predictions ({\sc Powheg-box}\ and {\sc MG5\_aMC}) are only $0.7\%$ apart. This simply means that the details 
    of the various VBS approximations have an impact below $1\%$ at the level of the fiducial cross section at LO for a 
    typical phase-space volume used by experimental collaborations. This is in agreement with the findings of 
    Refs.~\cite{Denner:2012dz,Oleari:2003tc}. Second, the four complete predictions ({\sc WHIZARD}, {\sc Phantom}, {\sc MG5\_aMC}, 
    and {\sc MoCaNLO+Recola}) are not in statistical agreement. While we have checked the point-wise agreement of the matrix-element, 
    we cannot exclude other reasons for the disagreement, for example a non-representative (\emph{i.e.}\ too-aggressive) 
    estimate of the Monte Carlo uncertainty or a non-perfect mapping of the six-body phase-space. However, the level of ambiguity is at the $0.5\%$ level, which 
    we deem satisfactory compared to the larger differences observed at NLO or when including matching to parton shower. 
    \begin{table}
        \centering
        \begin{tabular}{c|r@{ $\pm$ }l}
          Code  &  \multicolumn{2}{c}{$\sigma[\rm{fb}]$}  \\
            \hline
            \hline
            {\sc Bonsay}  &  $1.43636$ & $0.00002$ \\
            {\sc Powheg-Box}  &  $1.44092$ & $0.00009$ \\
            {\sc VBFNLO}  &  $1.43796$ & $0.00005$ \\
            {\sc Phantom} &  $1.4374\phantom{0}$ & $0.0006 $  \\
            {\sc WHIZARD} &  $1.4381\phantom{0}$ & $0.0002 $ \\
            {\sc MG5\_aMC}&  $1.4304\phantom{0}$ & $0.0007$ \\
            {\sc MoCaNLO+Recola}  &  $1.43476$ & $0.00009$ 
        \end{tabular}
        \caption{\label{tab:wg1_LOrates} Cross sections at LO accuracy and order $\mathcal{O}(\alpha^6)$.
        The complete $2\to 6$ matrix-element, without any approximation, is employed by {\sc Phantom},
        {\sc WHIZARD}, {\sc MG5\_aMC}, and {\sc MoCaNLO+Recola}. The predictions are obtained 
        in the fiducial region described in Sec.~\protect\ref{subsec:inputpar}.
        The uncertainties shown refer to the estimated statistical errors of the Monte Carlo integrations.}
    \end{table}

\begin{figure*}
       \centering
       \includegraphics[width=0.4\textwidth,angle=0,clip=true,trim={0.4cm 2cm 0.cm 1.cm}]{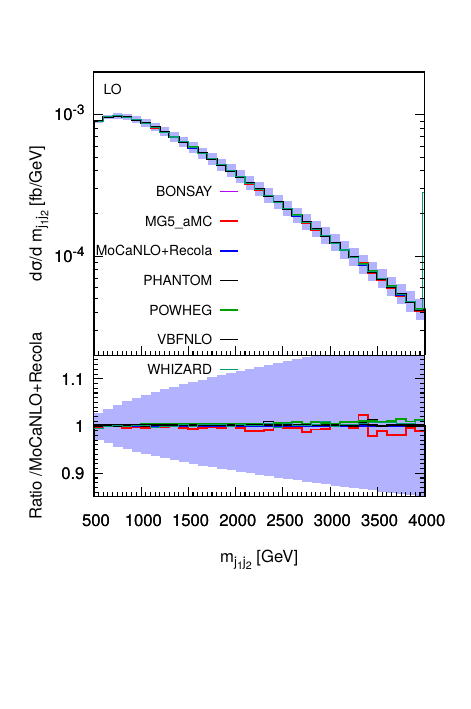}
       \includegraphics[width=0.4\textwidth,angle=0,clip=true,trim={0.3cm 2cm 0.cm 1.cm}]{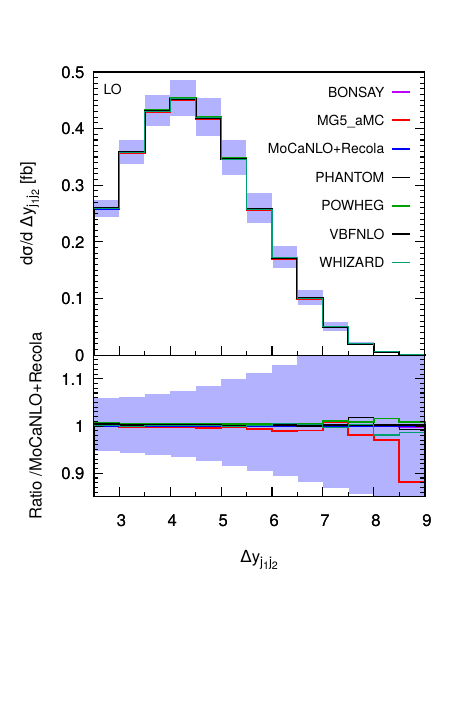}
    \caption{\label{fig:wg1_mjj-llLO} Differential distributions in the invariant mass (left) and rapidity difference of the two tagging jets (right) at LO accuracy \emph{i.e.}\ at order $\mathcal{O}(\alpha^6)$.
    The description of the different programs used can be found in Sec.~\protect\ref{subsec:codedescr}.
    The upper plots provides the absolute value for each prediction while the lower plots presents all predictions normalised to {\sc MoCaNLO}+{\sc Recola} which is one of the programs that provide the full prediction.
    The band corresponds to a seven-point variation of the renormalisation and factorisation scales.
    The predictions are obtained in the fiducial region described in Sec.~\protect\ref{subsec:inputpar}.}
    \end{figure*}

    In Fig.~\ref{fig:wg1_mjj-llLO}, we show the distributions in the invariant mass (left) and the rapidity difference (right) of the two tagging jets which are key observables for VBS measurements.
    In both cases we show the absolute distributions in the upper plot, while the lower plot displays the ratio over the predictions of {\sc MoCaNLO+Recola}, 
    for which we also display the scale-uncertainty band (seven-points variation as in Eq.~(3.11) of Ref.~\cite{Biedermann:2017bss}).
    For both observables we find a relatively good agreement among the various tools, which confirms the fact that contributions from $s$-channel diagrams as well as interferences are suppressed in the fiducial region.
    In general, the agreement is at the level of $1\%$ or below in each bin.
    We have checked that the same level of agreement holds for other standard differential distributions such as rapidity, invariant mass, or transverse momentum.
    This means that at LO, in the fiducial volume and for energies relevant to the LHC, the VBS approximation is good to a per cent.
    This is in agreement with the findings of Sec.~\ref{subsec:LOinclusive} as the present comparison completely excludes the phase-space region where tri-boson contributions could have a noticeable impact.

\section{Next-to-leading order QCD}
    \label{sec:NLO}
    \subsection{Inclusive comparison}
        \label{subsec:NLOinclusive}
    According to the results of Secs.~\ref{subsec:contributions} and \ref{subsec:LOinclusive}, the VBS approximation at LO fails drastically in the region $m_{\Pj\Pj} < 200$ GeV, $|\Delta y_{\Pj\Pj}| < 2$.
    Therefore, we present an inclusive study at NLO QCD for the EW component, namely the order $\mathcal{O}(\alphas\alpha^6)$ for the set-up described in Sec.~\ref{subsec:inputpar} but imposing the requirements $m_{\Pj\Pj}>200 \GeV$ and $|\Delta y_{\Pj\Pj}|>2$.

    We compare three different predictions at NLO QCD: 
    the VBS approximation implemented in {\sc Bonsay} (dubbed $|t|^2+|u|^2$), the VBS approximation with the $s$-channel contributions from {\sc VBFNLO} (dubbed $|s|^2+|t|^2+|u|^2$), and the full computation.
    The full computation employs exact matrix elements meaning that $t/u/s$ interferences, factorisable and non-factorisable QCD corrections, as well as EW corrections to the order $\mathcal{O}(\alphas \alpha^5)$ are included.

    The total cross sections within the above-mentioned kinematic cuts are shown in Tab.~\ref{tab:crosssecINCLUSIVE}.
    The $|t|^2+|u|^2$ approximation for NLO QCD predictions is lower by about $6\%$ than the full calculation.
    The inclusion of $s$-channel diagrams improves the approximate prediction, leading to an excess at the $3\%$ level.

    \begin{table}
    \centering
    \begin{tabular}{c|c|c}
    Prediction & $\sigma_{\textrm{tot}}\,[\textrm{fb}]$ & $\delta [\%]$ \\
    \hline
    \hline
    full &  $1.733\phantom{0} \pm 0.002\phantom{0}$ & - \\
    \hline
    $|t|^2 + |u|^2$ & $1.6292 \pm 0.0001$  &  $-6.0$ \\
    \hline
    $|s|^2 + |t|^2 + |u|^2$ & $1.7780 \pm 0.0001$  & $+2.6$
    \end{tabular}
    \caption{Cross sections at NLO QCD \emph{i.e.}\ at order $\mathcal{O}(\alphas\alpha^6)$ for the full computation and two approximations.
    In addition to the cuts of Sec.~\protect\ref{subsec:inputpar}, the VBS cuts take the values $m_{\Pj\Pj}>200 \GeV$ and $|\Delta y_{\Pj\Pj}|>2$.
    The uncertainties shown refer to the estimated statistical errors of the Monte Carlo programs.}
    \label{tab:crosssecINCLUSIVE}
    \end{table}

\begin{figure*}
    \centering
    {\includegraphics[scale=0.35]{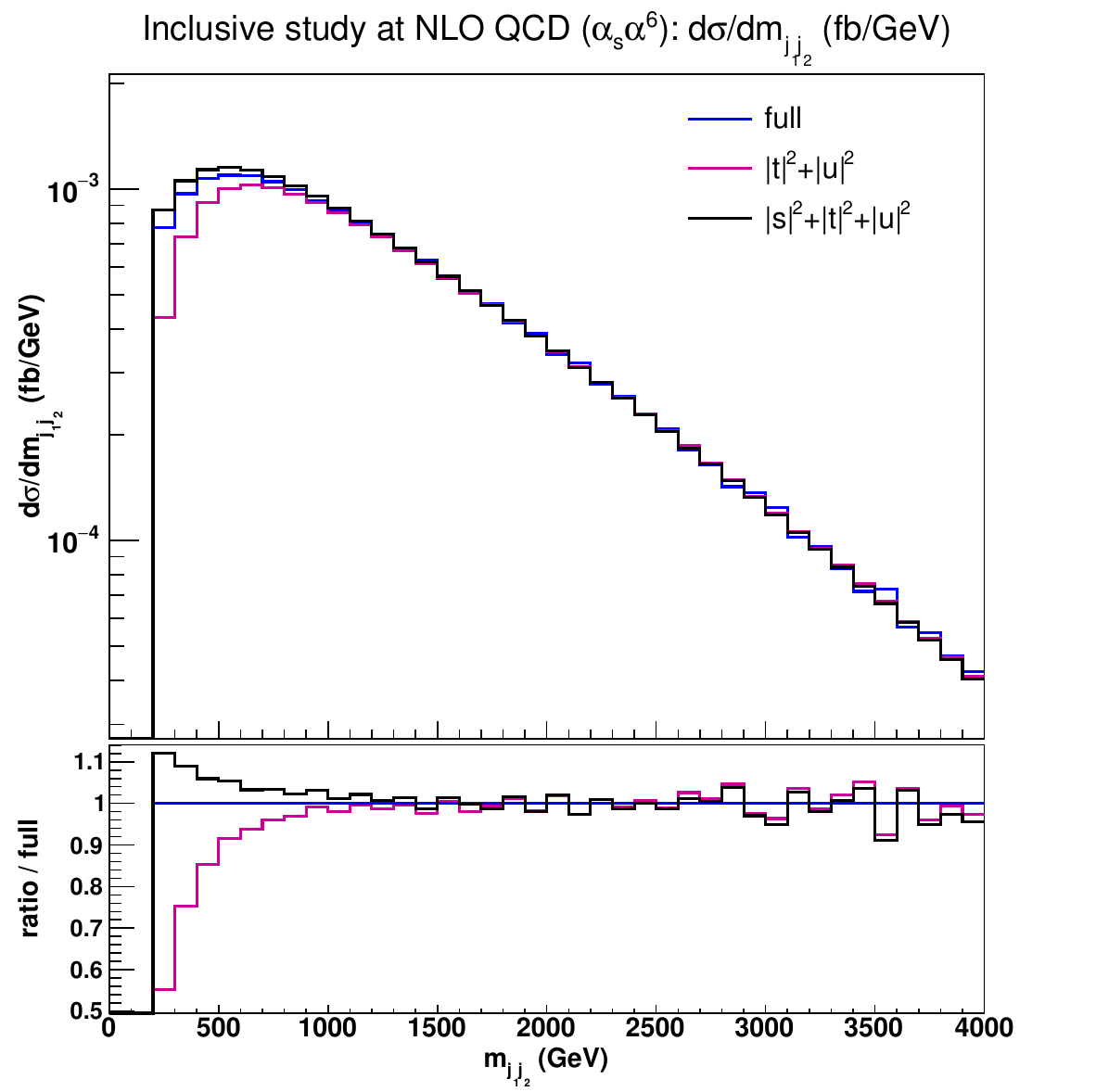}}
    {\includegraphics[scale=0.35]{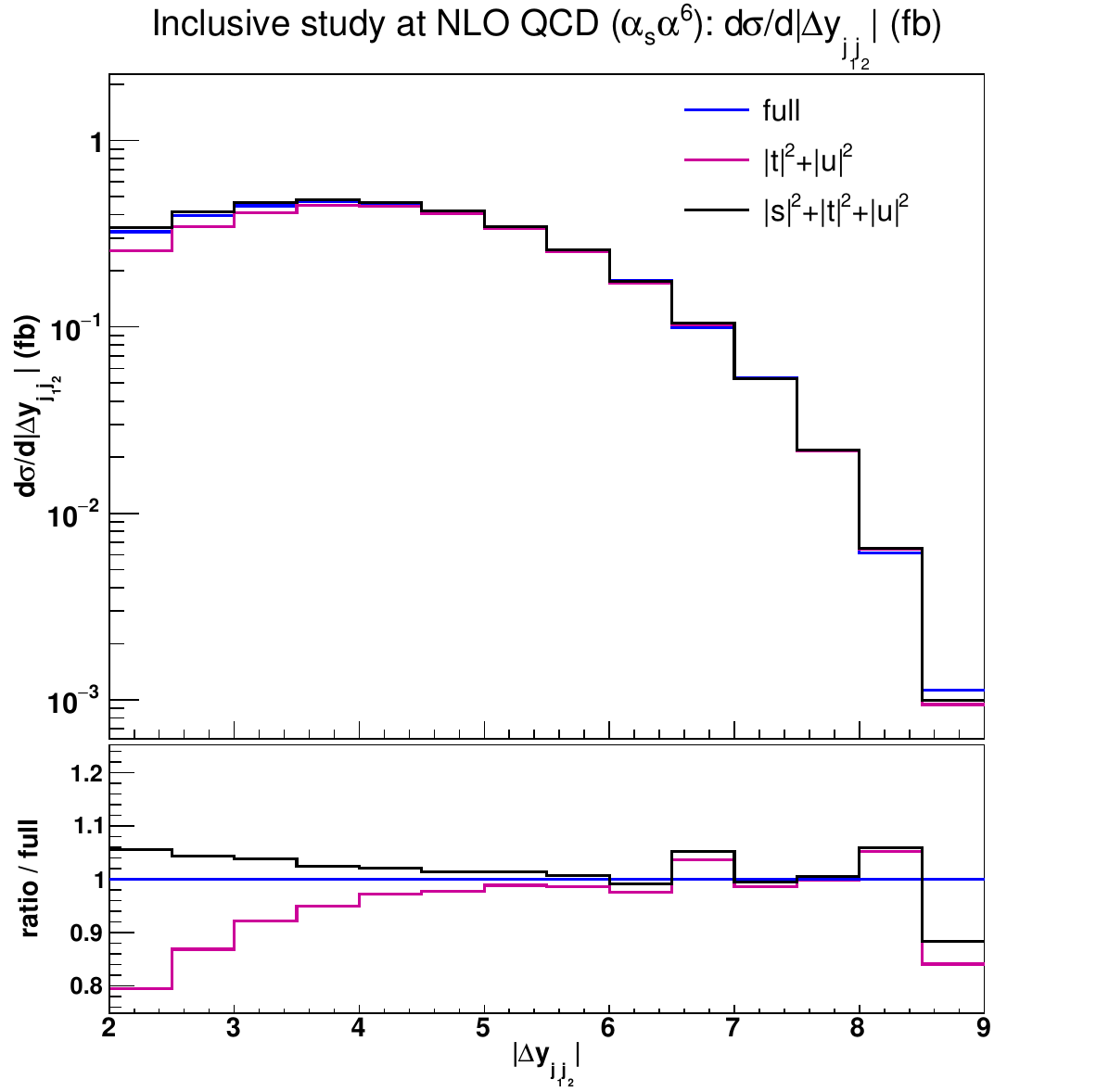}}
    \caption{
    Differential distributions in the di-jet invariant mass (left) and the rapidity separation (right) of the two tagging jets at NLO QCD \emph{i.e.}\ at order $\mathcal{O}(\alphas\alpha^6)$ for the full computation and two approximations.
    The upper plots provide the absolute value for each prediction while the lower plots present all predictions normalised to {\sc MoCaNLO}+{\sc Recola} which is one of the programs that provide the full prediction.
    In addition to the cuts of Sec.~\protect\ref{subsec:inputpar}, the VBS cuts take the values $m_{\Pj\Pj}>200 \GeV$ and $|\Delta y_{\Pj\Pj}|>2$.} 
    \label{fig:mjjdyjj_1d_1}
    \end{figure*}

    These differences are more evident in differential distributions.
    In Fig.~\ref{fig:mjjdyjj_1d_1}, we show the differential distributions in the di-jet invariant mass $m_{\Pj\Pj}$ and rapidity separation $|\Delta y_{\Pj\Pj}|$.
    For large $m_{\Pj\Pj}$ and large $|\Delta y_{\Pj\Pj}|$, as expected, the VBS approximation is performing well and its $s$-channel extension agrees with the full calculation within $10\%$.
    This is in contrast with the regions  $200 \GeV < m_{\Pj\Pj} < 500 \GeV$ and $2<|\Delta y_{\Pj\Pj}|<2.5$, where
    the difference between the $|t|^2+|u|^2$ approximation and the full computation can be above $30\%$.
    The inclusion of $s$-channel contributions cures partly this behaviour by improving the approximation to about $10\%$.
    This tends to indicate that interference contributions and/or non-factorisable QCD corrections play a non-negligible role in this phase-space region.

\begin{figure*}
    \centering
    {\includegraphics[scale=0.395]{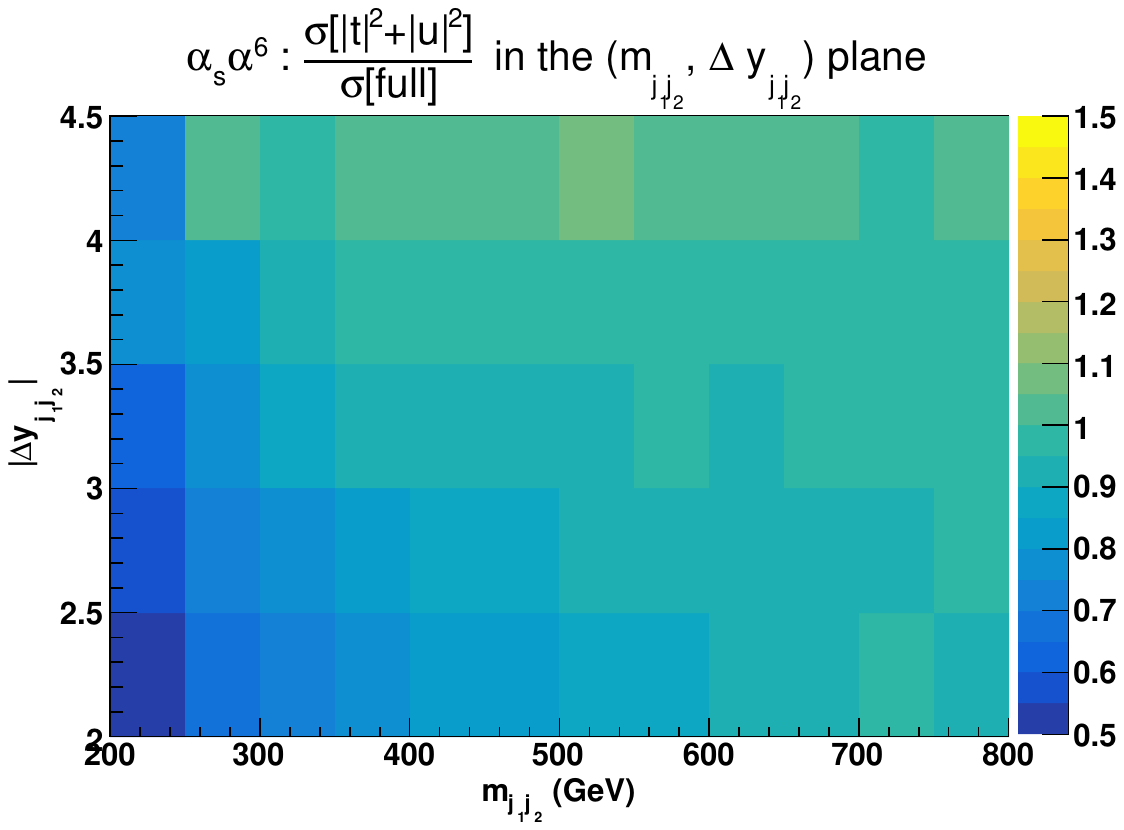}}
    {\includegraphics[scale=0.395]{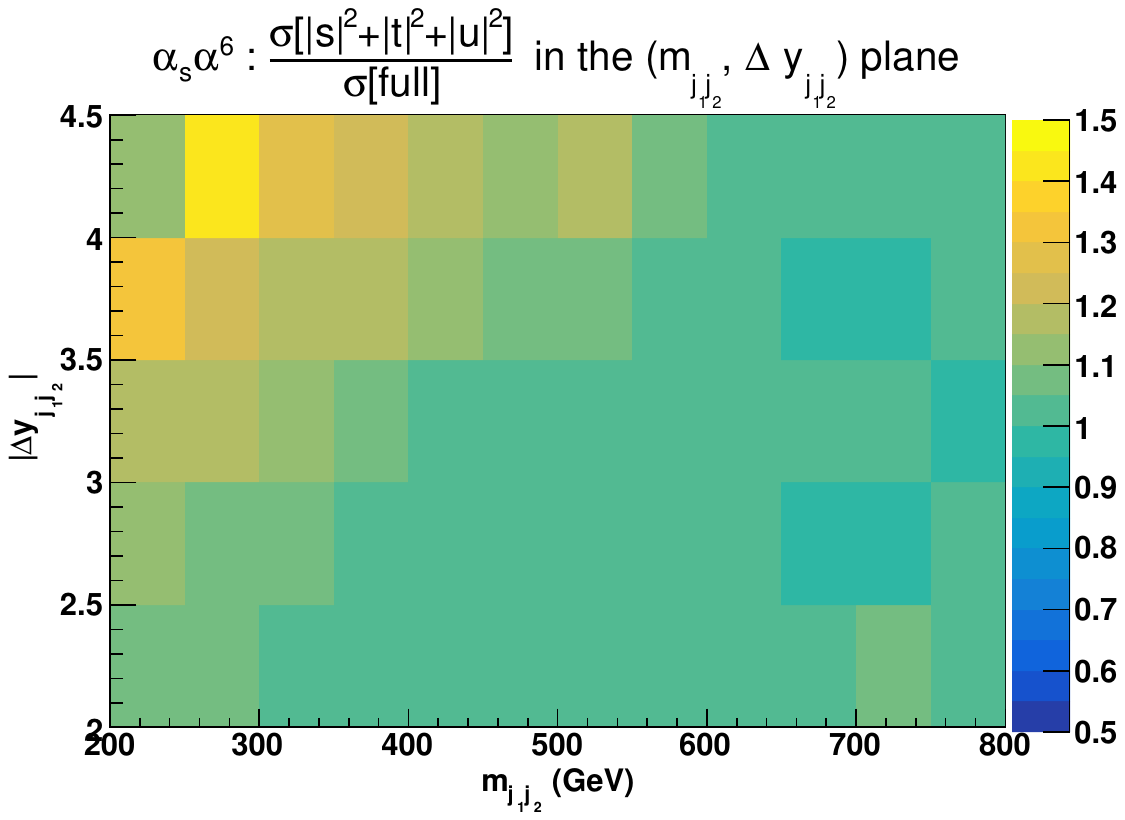}}
    \caption{Ratios for double-differential distributions in the variables $m_{\Pj\Pj}$ and $|\Delta y_{\Pj\Pj}|$ at NLO QCD \emph{i.e.}\ at order $\mathcal{O}(\alphas\alpha^6)$ of the approximated squared amplitudes over the full matrix element.
    The approximated squared amplitudes are computed as $|\mathcal{A}|^2 \sim |t|^2+|u|^2$ (left) and $|\mathcal{A}|^2 \sim |s|^2+|t|^2+|u|^2$ (right).
    In addition to the cuts of Sec.~\protect\ref{subsec:inputpar}, the VBS cuts take the values $m_{\Pj\Pj}>200 \GeV$ and $|\Delta y_{\Pj\Pj}|>2$.}
    \label{fig:ratio2d_NLO}
    \end{figure*}

    In order to investigate further the jet-pair kinematics, we study the double-differential distribution in the variables $m_{\Pj\Pj}$ and $|\Delta y_{\Pj\Pj}|$.
    In particular, in Fig.~\ref{fig:ratio2d_NLO}, we compute in each bin the ratios of the approximated cross sections over the full ones [$\sigma(|t|^2+|u|^2)/\sigma(\textrm{full})$ and $\sigma(|s|^2+|t|^2+|u|^2)/\sigma(\textrm{full})$].
    As expected, in the low invariant-mass and low rapidity-separation region of the jet pair ($200 \GeV < m_{\Pj\Pj} < 500 \GeV$, $2<|\Delta y_{\Pj\Pj}|<2.5$) the VBS approximation fails significantly (by more than $40\%$).
    Including the $s$-channel contributions leads to a difference of less than $10\%$ in this very region.
    However, in the region of large di-jet invariant mass and low rapidity separation of the jets, the $|s|^2+|t|^2+|u|^2$ approximation overestimates the full computation by more than $40\%$.\footnote{The bin
    in the top-left corner of the right-hand-side plot of Fig.~\ref{fig:ratio2d_NLO} suffers from large uncertainty ($30\%$) while the other errors are at the per-cent level.}
    Again, this seems to support the fact that interferences and non-factorisable corrections can be non-negligible in this region.
    On the other hand, in the typical VBS region, the VBS approximation shows a good agreement with the full computation as documented in detail in Sec.~\ref{sec:fidNLO}.

\begin{figure*}
    \centering
    {\includegraphics[scale=0.35]{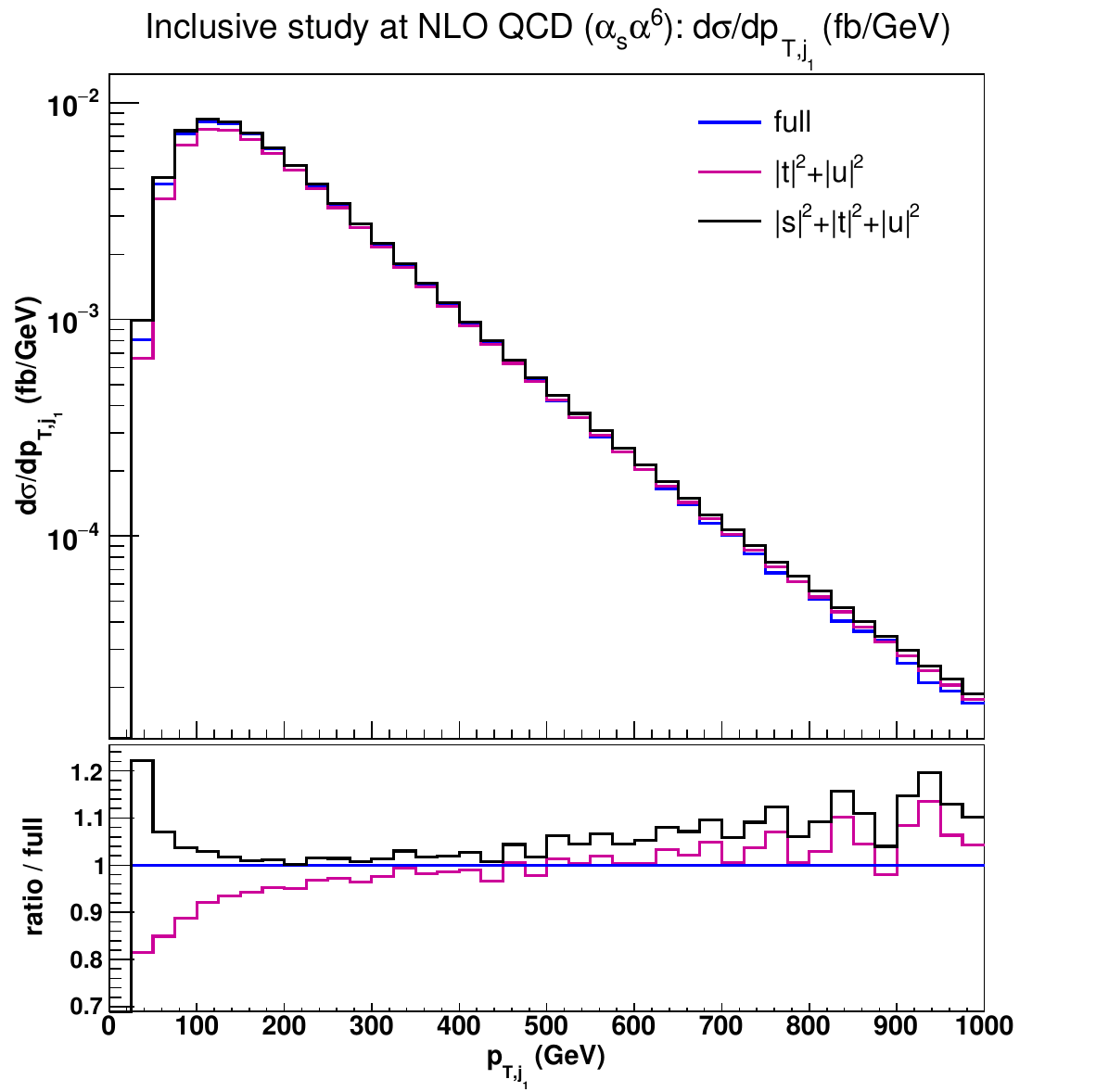}}
    {\includegraphics[scale=0.35]{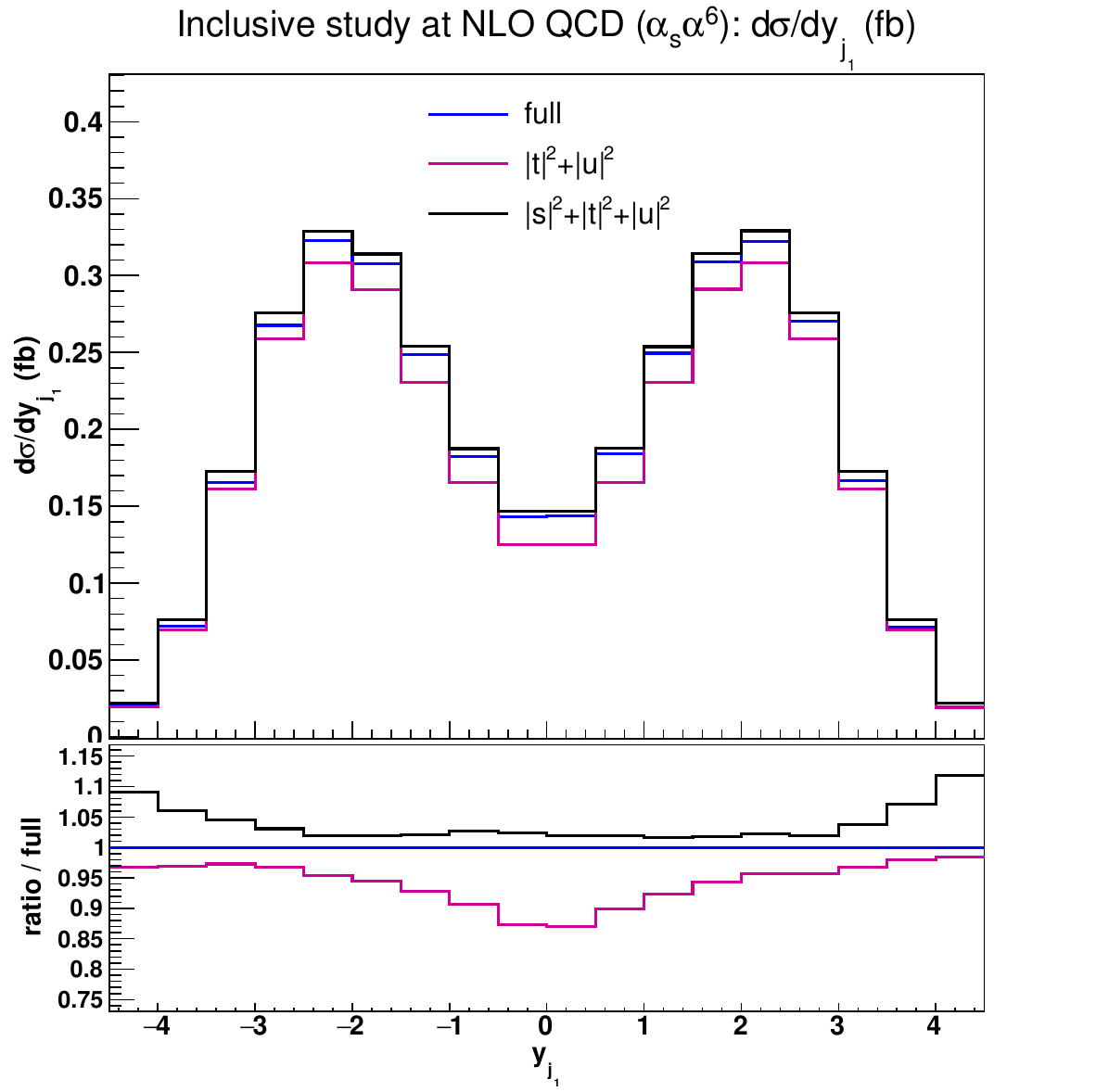}}
    \caption{Differential distributions in the transverse momentum (left) and rapidity (right) of the hardest tagging jet at NLO QCD \emph{i.e.}\ at order $\mathcal{O}(\alphas\alpha^6)$ for the full computation and two approximations.
    The upper plots provide the absolute value for each prediction while the lower plots present all predictions normalised to {\sc MoCaNLO}+{\sc Recola} which is one of the programs that provide the full prediction.
    In addition to the cuts of Sec.~\protect\ref{subsec:inputpar}, the VBS cuts take the values $m_{\Pj\Pj}>200 \GeV$ and $|\Delta y_{\Pj\Pj}|>2$.} 
    \label{fig:mjjdyjj_1d_2}
    \end{figure*}

    In Fig.~\ref{fig:mjjdyjj_1d_2}, the distributions in the transverse momentum of the hardest jet and its rapidity are shown.
    At low transverse momentum, $|t|^2+|u|^2$ and $|s|^2+|t|^2+|u|^2$ approximations are lower and higher than the full computation by about $20\%$, respectively.
    At high transverse momentum, they have a similar behaviour.
    They both diverge from the full computation towards larger transverse momentum (about $10\%$ at $1000\GeV$).
    Regarding the rapidity of the hardest jet, the two approximations have opposite behaviours.
    In the central region, the $|t|^2+|u|^2$ approximation differs by $12\%$ with respect to the full computation, while the $|s|^2+|t|^2+|u|^2$ one is good within $5\%$.
    In the peripheral region, the $|t|^2+|u|^2$ approximation is rather close to the full computation ($5\%$), while the $|s|^2+|t|^2+|u|^2$ one differs by $10\%$.

\begin{figure*}
    \centering
    {\includegraphics[scale=0.35]{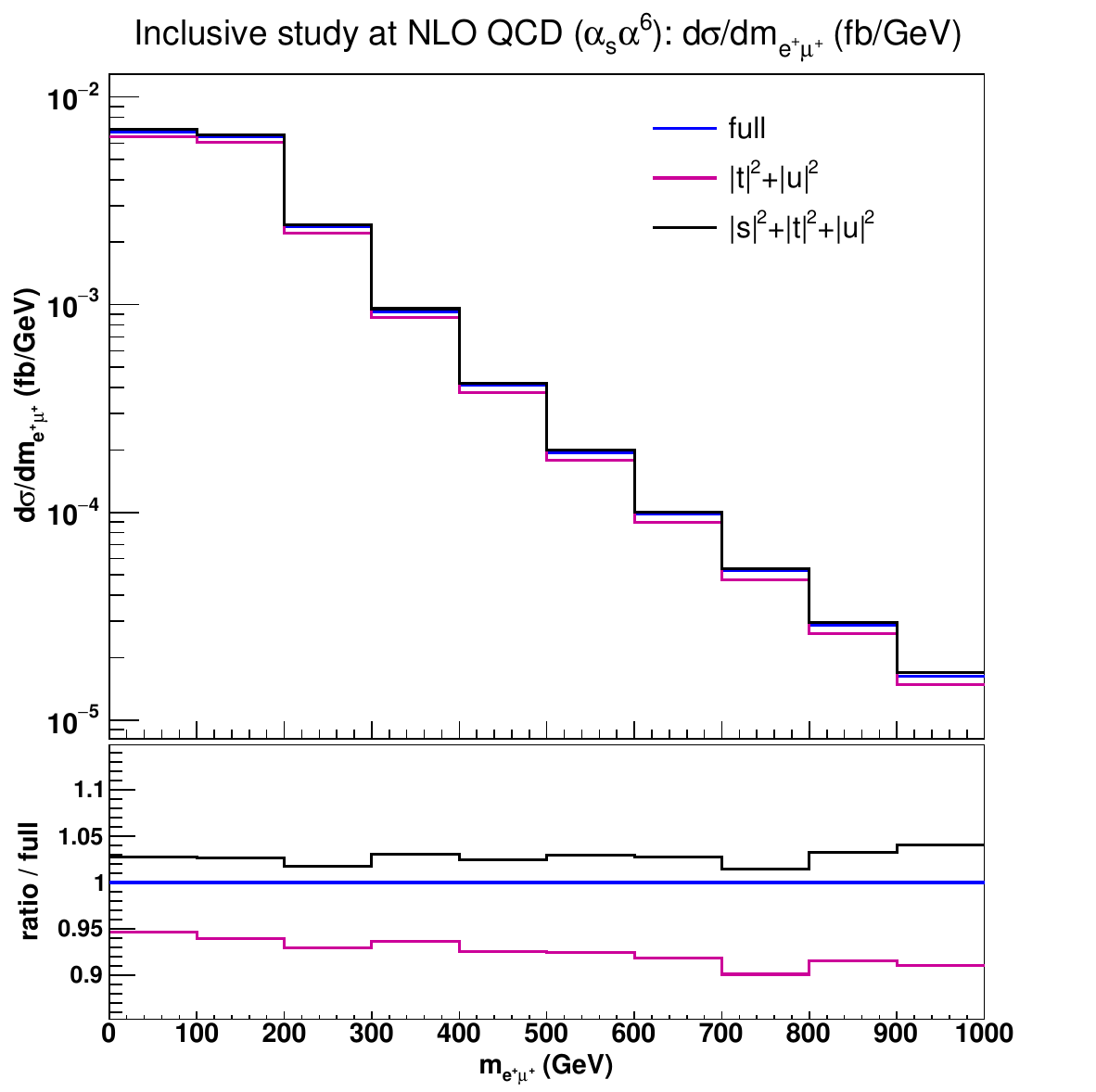}}
    {\includegraphics[scale=0.35]{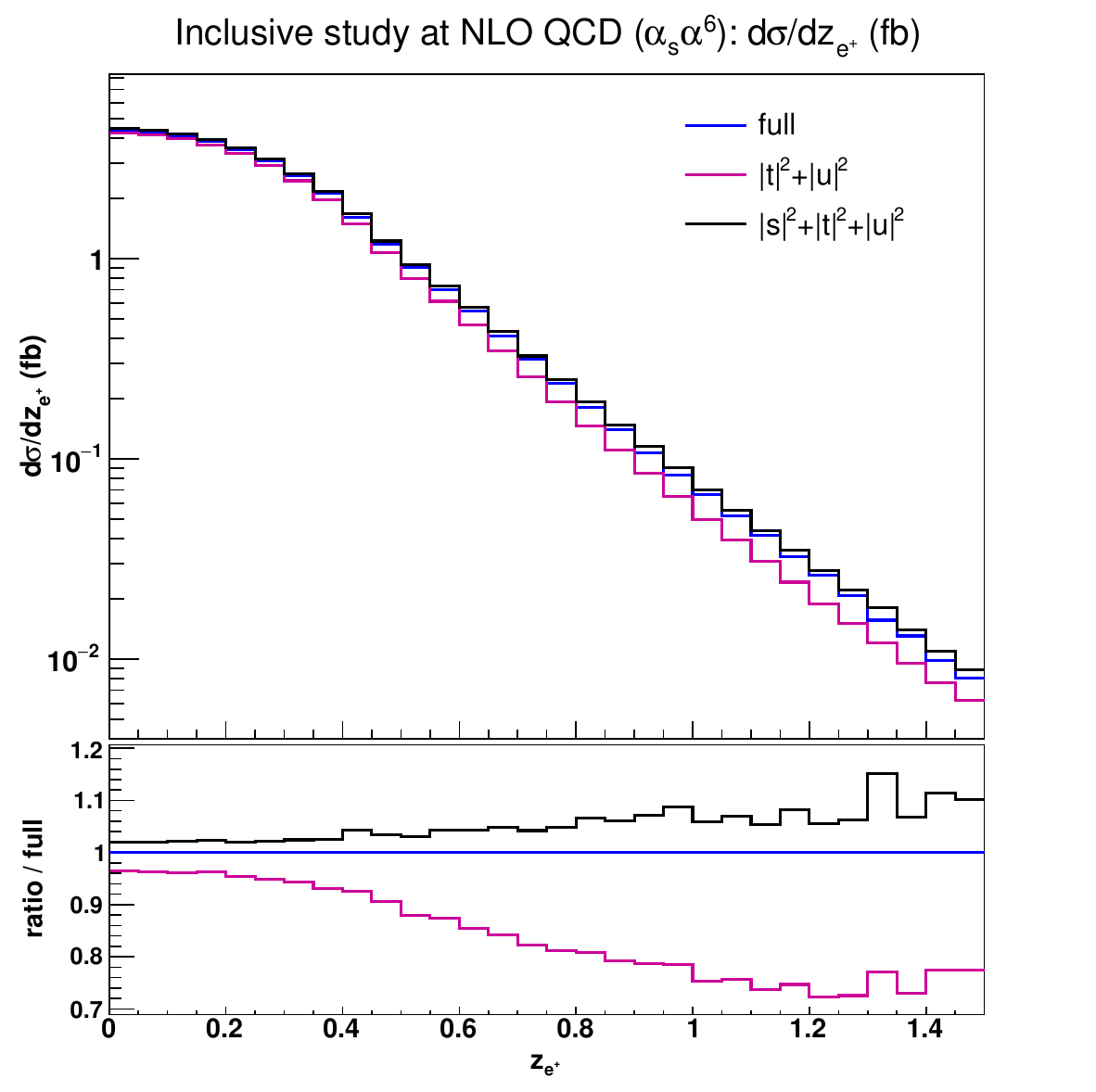}}
    \caption{Differential distributions in the lepton-lepton invariant mass (left) and the electron Zeppenfeld variable (right) at NLO QCD \emph{i.e.}\ at order $\mathcal{O}(\alphas\alpha^6)$ for the full computation and two approximations.
    The upper plots provide the absolute value for each prediction while the lower plots presents all predictions normalised to {\sc MoCaNLO}+{\sc Recola} which is one of the programs that provide the full prediction.
    In addition to the cuts of Sec.~\protect\ref{subsec:inputpar}, the VBS cuts take the values $m_{\Pj\Pj}>200 \GeV$ and $|\Delta y_{\Pj\Pj}|>2$.} 
    \label{fig:mjjdyjj_1d_3}
    \end{figure*}

    Concerning leptonic observables, we show in Fig.~\ref{fig:mjjdyjj_1d_3} the distributions in the di-lepton invariant mass and in the Zeppenfeld variable of the electron, defined as
    \begin{equation}
      z_{\Pe^+} = \frac{y_{\Pe^+}-\frac{y_{\Pj_1}+y_{\Pj_2}}2}{|\Delta y_{jj}|} \,.
      \label{eq:Zeppenfeld}
    \end{equation}
    Analogous definitions are later also used for the Zeppenfeld variable of the muon and of the third jet.
    The $|s|^2+|t|^2+|u|^2$ predictions for $m_{\rm e^+\mu^+}$ agree rather well with the full curve, obtained from {\sc MoCa\-NLO+Recola}.
    The prediction from {\sc Bonsay} is about $10\%$ lower around $1000 \GeV$.
    The Zeppenfeld variable of the positron $z_e$ is more strongly affected by the exclusion of $s$-channel contributions.
    For increasing $z_{\rm e}$, the $|t|^2+|u|^2$ approximation diverges from the full computation to reach a difference of about $25\%$ at $1.5$.
    On the other hand, including $s$-channel contributions leads to a better approximation, staying within $10\%$ difference over the whole range.

    In conclusion, both the loose minimum di-jet invariant-mass cut and the inclusion of QCD radiative corrections render the $s$-channel contributions less suppressed than at LO, making their inclusion mandatory, in order to provide trustworthy predictions at NLO accuracy.
    In the inclusive region studied here, neglecting $s$-channel contributions, non-factorisable corrections, and EW corrections can lead to discrepancies of up to $30\%$ with respect to the full computation.
    Nevertheless, the VBS approximation at NLO provides a good approximation of full calculations in the kinematic region where $m_{\Pj\Pj} \gtrsim 500 \GeV$ and $|\Delta y_{\Pj\Pj}| \gtrsim 2.5$), for both total cross section and differential distributions.
    This more exclusive region is studied in more detail in the next section.
    \subsection{Comparison in the fiducial region}
    \label{sec:fidNLO}
    In Tab.~\ref{tab:wg1_NLOrates}, the cross sections of the various tools at NLO-QCD accuracy are presented.
    The order considered is again the order $\mathcal{O}(\alpha_{\rm s}\alpha^6)$, and the fiducial volume is the one described in Sec.~\ref{subsec:inputpar}.
    In contrast with Tab.~\ref{tab:wg1_LOrates}, the NLO predictions differ visibly according to the approximations used.

    \begin{table}
        \centering
        \begin{tabular}{c|r@{ $\pm$ }l}
          Code  &  \multicolumn{2}{c}{$\sigma[\rm{fb}]$}  \\
            \hline
            \hline
            {\sc Bonsay}  &  $1.35039$ & $0.00006$  \\
            {\sc Powheg-Box}  &  $1.3605\phantom{0}$  & $0.0007$   \\
            {\sc VBFNLO}  &  $1.3916\phantom{0}$ & $0.0001$  \\
            {\sc MG5\_aMC}&  $1.363\phantom{0}\phantom{0}$ & $0.004$  \\
            {\sc MoCaNLO+Recola}  &  $ 1.378\phantom{0}\phantom{0}$ & $0.001$ 
        \end{tabular}
        \caption{\label{tab:wg1_NLOrates} Cross sections at NLO accuracy and order $\mathcal{O}(\alphas\alpha^6)$.
        The predictions are obtained in the fiducial region described in Sec.~\protect\ref{subsec:inputpar}.
        The uncertainties shown refer to estimated statistical errors of the Monte Carlo integrations.}
    \end{table}

    The first observation is that the predictions featuring two versions
    of the VBS approximation ({\sc Bonsay} and the {\sc Powheg-Box}) are
    relatively close.\footnote{The {\sc VBFNLO} prediction omitting
    $s$-channel contributions amounts to $1.3703 \pm 0.0001$ fb. This differs from
    the {\sc Powheg-Box} prediction mainly due to the different choice of
    scales used in the {\sc Powheg-Box} (\emph{cf.}
    footnote \ref{foot:powheg}).} {\sc Bonsay} uses a double-pole
    approximation for the virtual matrix element, and it is worth noticing
    that this approximation seems to be accurate at $1\%$ level as
    compared to the {\sc Powheg-Box}. This means that the double-pole
    approximation on the two W bosons used in {\sc Bonsay} constitutes a
    good approximation of the VBS-approxi\-mated virtual corrections
    implemented in the {\sc Powheg-Box}.  Both predictions differ by about
    $2\%$ with respect to the full computation ({\sc MoCaNLO+Recola}).
    The second observation is that the inclusion of $s$-channel
    contributions seems to have a significant impact.  Indeed, their
    inclusion (as done in {\sc VBFNLO}) approximates the full computation
    by a per cent. The main contribution due to 
    the $s$-channel diagrams thereby consists of real-emission
    contributions, where one of the two leading jets is formed by one
    quark, or possibly also both quarks, originating from the W-boson decay,
    and the second one by the extra radiation emitted from the initial
    state. In such configurations, the hadronically-decaying W boson can
    become on-shell and hence yield larger contributions than at LO, where
    the invariant mass cut on the two jets forces the boson into the far
    off-shell region.
    However, the agreement between {\sc MoCaNLO+Recola} and {\sc VBFNLO} is mostly accidental, as the inclusion of interference effects and some non-factorisable corrections (in the real corrections) in {\sc MG5\_aMC} brings the prediction down and closer to the VBS approximation.
    Not unexpectedly none of the approximations used here agrees perfectly with the full calculation of {\sc MoCaNLO\-+Recola} at NLO.
    Nevertheless, the disagreement seems never to exceed $2\%$ at the fiducial cross-section level.

\begin{figure*}
       \centering
       \includegraphics[width=0.4\textwidth,angle=0,clip=true,trim={0.4cm 2cm 0.cm 1.cm}]{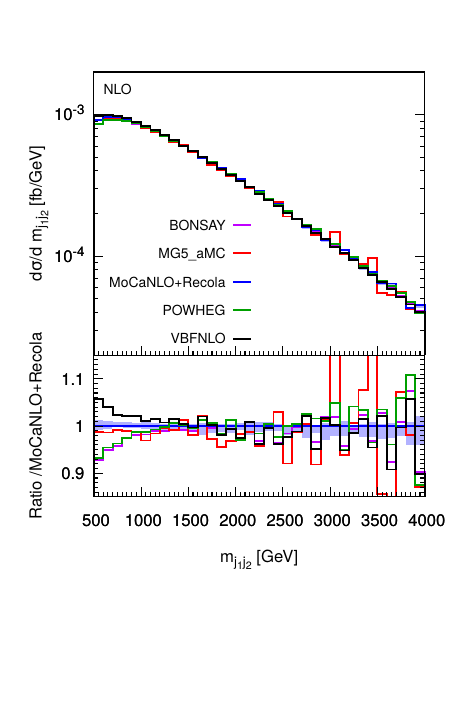}
       \includegraphics[width=0.4\textwidth,angle=0,clip=true,trim={0.4cm 2cm 0.cm 1.cm}]{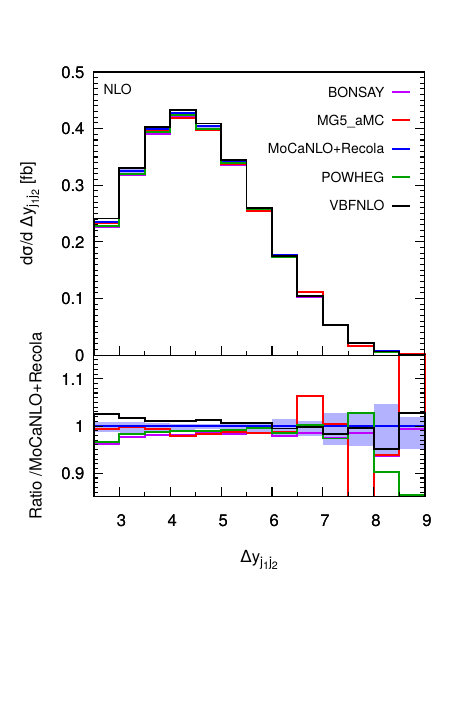}
    \caption{\label{fig:distNLO1} Differential distributions in the invariant mass (left) and rapidity difference (right) of the two tagging jets at NLO accuracy \emph{i.e.}\ at order $\mathcal{O}(\alpha_{\rm s}\alpha^6)$.
    The description of the different programs used can be found in Sec.~\protect\ref{subsec:codedescr}.
    The upper plots provide the absolute value for each prediction while the lower plots present all predictions normalised to {\sc MoCaNLO}+{\sc Recola} which is the full prediction.
    The band corresponds to a seven-point variation of the renormalisation and factorisation scales.
    The predictions are obtained in the fiducial region described in Sec.~\protect\ref{subsec:inputpar}.
    }
    \end{figure*}

\begin{figure*}
       \centering
       \includegraphics[width=0.4\textwidth,angle=0,clip=true,trim={0.4cm 2cm 0.cm 1.cm}]{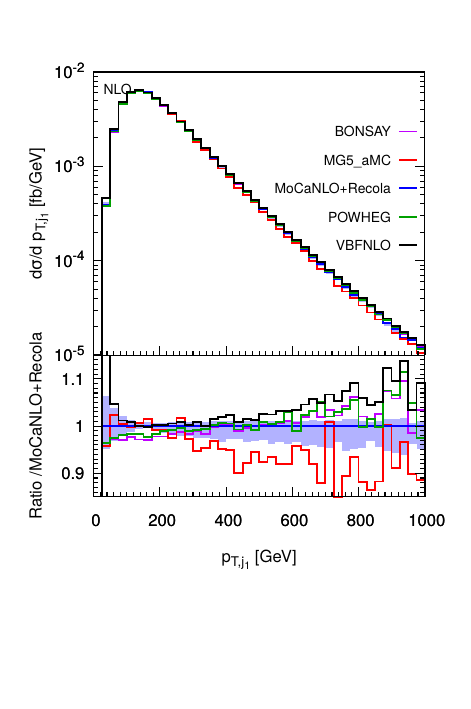}
       \includegraphics[width=0.4\textwidth,angle=0,clip=true,trim={0.3cm 2cm 0.cm 1.cm}]{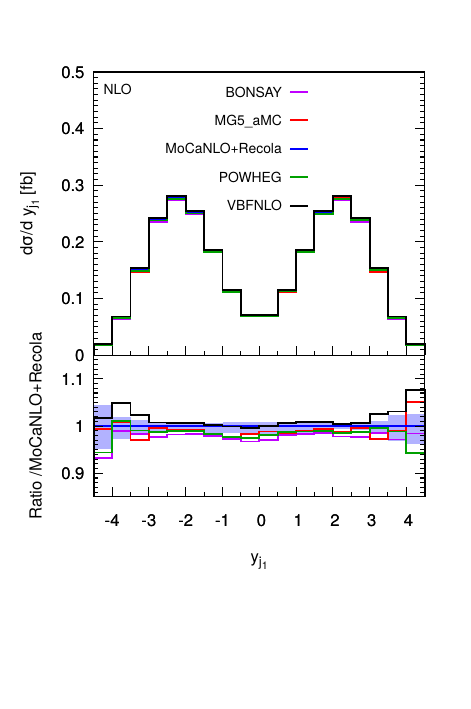}
    \caption{\label{fig:distNLO2} Differential distributions in the transverse momentum (left) and rapidity (right) of the hardest jet at NLO accuracy \emph{i.e.}\ at order $\mathcal{O}(\alpha_{\rm s}\alpha^6)$.
    The description of the different programs used can be found in Sec.~\protect\ref{subsec:codedescr}.
    The upper plots provide the absolute value for each prediction while the lower plots present all predictions normalised to {\sc MoCaNLO}+{\sc Recola} which is the full prediction.
    The band corresponds to a seven-point variation of the renormalisation and factorisation scales.
    The predictions are obtained in the fiducial region described in Sec.~\ref{subsec:inputpar}.
    }
    \end{figure*}

\begin{figure*}
       \centering
       \includegraphics[width=0.4\textwidth,angle=0,clip=true,trim={0.4cm 2cm 0.cm 1.cm}]{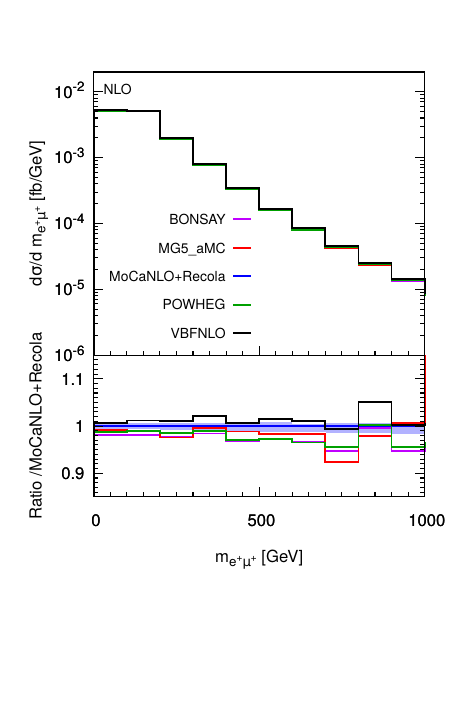}
       \includegraphics[width=0.4\textwidth,angle=0,clip=true,trim={0.4cm 2cm 0.cm 1.cm}]{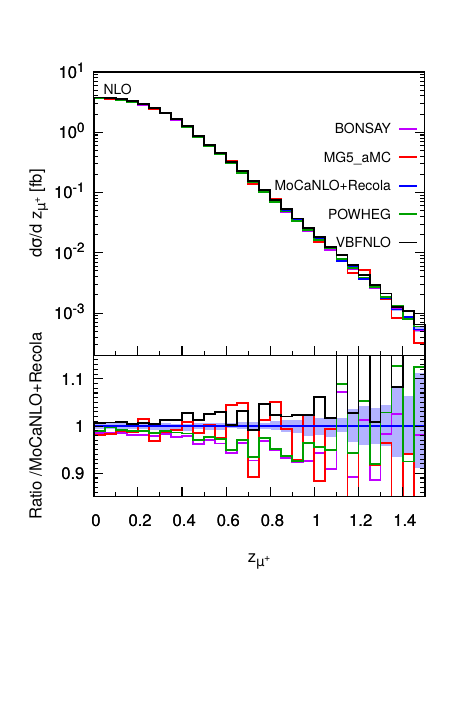}
    \caption{\label{fig:distNLO3} Differential distributions in the invariant mass of the two charged leptons (left) and Zeppenfeld variable for the muon (right) at NLO accuracy \emph{i.e.}\ at order $\mathcal{O}(\alpha_{\rm s}\alpha^6)$.
    The description of the different programs used can be found in Sec.~\protect\ref{subsec:codedescr}.
    The upper plots provide the absolute value for each prediction while the lower plots present all predictions normalised to {\sc MoCaNLO}+{\sc Recola} which is the full prediction.
    The band corresponds to a seven-point variation of the renormalisation and factorisation scales.
    The predictions are obtained in the fiducial region described in Sec.~\protect\ref{subsec:inputpar}.
    }
    \end{figure*}

    In Figs.~\ref{fig:distNLO1}--\ref{fig:distNLO3}, several differential distributions are shown.
    All these predictions are performed at NLO accuracy at the order $\mathcal{O}(\alphas\alpha^6)$.
    In the upper panel, the absolute predictions are shown while in the lower panel, the ratio with respect to the full predictions are displayed.
    The band corresponds to a seven-points variation of the factorisation and renormalisation scales (as defined in Eq.~(3.11) of Ref.~\cite{Biedermann:2017bss}).

    We start with Fig.~\ref{fig:distNLO1} which displays the invariant mass (left) and the rapidity separation (right) of the two tagging jets.
    For high invariant mass, all predictions agree rather well.
    On the other hand, for low invariant mass, the hierarchy present at the level of the cross section is reproduced.
    The VBS-approximated predictions ({\sc Bonsay} and {\sc Powheg-Box}) are lower than the full calculation ({\sc MoCaNLO}+{\sc Recola}).
    The full calculation is rather well approximated by the hybrid VBS approximation implemented in {\sc MG5\_aMC}.
    Finally, {\sc VBFNLO} which includes also $s$-channel contributions provides larger predictions at low invariant mass.
    For the rapidity difference between the two tagging jets, the hierarchy between the predictions is rather similar.
    Therefore, depending on the approximation used, it can vary by $\pm7\%$ and $\pm4\%$ with respect to the full computation at low invariant mass and low rapidity difference for the tagging jets, respectively

    Concerning the transverse momentum (left) and rapidity (right) of the hardest jet shown in Fig.~\ref{fig:distNLO2}, the situation is rather different.
    While {\sc MG5\_aMC} is very close to the full prediction for low transverse momentum, it departs from it 
    at larger transverse momentum by about $10\%$.
    This is in contrast with the VBS-approximated predictions such as {\sc Bonsay}, {\sc Powheg}, and {\sc VBFNLO} which are lower than the full computation at low transverse momentum and higher for larger transverse momentum.
    The difference at high transverse momentum between the latter predictions and the full computation can be attributed to EW Sudakov logarithms that become large in this phase-space region.
    While the predictions of {\sc Bonsay} and {\sc Powheg} are rather close over the whole range, the one of {\sc VBFNLO} is very different at low transverse momentum where it is even higher than the full computation.
    We note that for the transverse momentum of the second hardest jet, the predictions from {\sc MG5\_aMC} are in good agreement with the other VBS-approximated predictions.
    Concerning the rapidity of the hardest jet, {\sc VBFNLO} is in good agreement with {\sc MoCaNLO}+{\sc Recola} in the rapidity range $|y_{j_1}| < 3$.
    For larger rapidity, the other codes constitute a better description of the full process at order $\mathcal{O}(\alpha_{\rm s}\alpha^6)$.

    The last set of differential distributions is the invariant mass of the two charged leptons (left) and the Zeppenfeld variable for the anti-muon (right).
    Concerning the comparison of the predictions, both distributions display a rather similar behaviour.
    Indeed, the hierarchy mentioned previously is here respected and enhanced towards high invariant mass or high Zeppenfeld variable.
    The predictions of {\sc MoCaNLO}+{\sc Recola} and {\sc VBFNLO} are in rather good agreement for both distributions for the kinematic range displayed here.
    The other three VBS approximations are close to each other within few per cent.

    In the end, the quality of the VBS approximations is good up to $10\%$ in the fiducial region.
    These differences are larger than those at LO.

    The contributions from the $s$-channel amplitude can be sizeable especially at low invariant mass for the two tagging jets (comparing the predictions of {\sc VBFNLO} against the ones of {\sc Bonsay} and {\sc Powheg}).
    This can be explained by the fact that $s$-channel contributions are less suppressed at NLO.
    As real radiation, an extra gluon-jet can be radiated from any of the strongly-interacting particles while the two quarks originating from the W-boson decay can be recombined in a single jet.
    Therefore, the jet requirements ($ m_{\Pj \Pj} >  500\GeV$ and $|\Delta y_{\Pj \Pj}| > 2.5$) that were suppressing $s$-channel contributions at LO are partially lifted with the inclusion of a third jet at NLO.
    Such an effect has also been observed for top--antitop production in the lepton+jet channel at NLO QCD \cite{Denner:2017kzu}.

    In phase-space regions where the $s$-channel contributions are sizeable their interference with the $t/u$-channel can be of similar size.
    This can be observed by comparing the predictions of {\sc VBFNLO} against the ones of {\sc MG5\_aMC}.

    Finally, the effect of EW corrections and non-\-factoris\hyp{}able contributions in the virtual corrections are usually small.
    But they can be relatively large (about $10\%$) for large transverse momentum of the hardest jet.
    These high-energy region of the phase space are where EW Sudakov logarithms become large.
    Nonetheless these regions are rather suppressed and thus these effects are hardly visible at the level of the cross section.

\section{Matching to parton shower}
    \label{sec:matching}
We now discuss how different predictions compare when the matching to parton shower is included, both at LO 
(\emph{i.e.}\ at order $\mathcal O (\alpha^6)$) and at NLO-QCD (\emph{i.e.}\ at order $\mathcal O (\alpha^6\alphas)$) accuracy. For such
a comparison we expect larger discrepancies than what we found at fixed order, as a consequence of the different
matching schemes, parton showers employed, and of other details of the matching (such as the choice of the parton shower initial scale). Among
the codes capable of providing fixed-order results, presented before, {\sc MG5\_aMC}, the {\sc Powheg-Box}, and {\sc VBFNLO}
can also provide results at (N)LO+PS accuracy. For {\sc VBFNLO} matched to {\sc Herwig} and the {\sc Powheg-Box}, we
restrict ourselves to show results only in the VBS approximation,
{\emph i.e.}\ the $s$-channel contributions are neglected here. Besides,
also {\sc Phantom} and {\sc WHIZARD} are used for LO+PS results.\\
{\sc MG5\_aMC}, which employs the {\sc MC@NLO}~\cite{Frixione:2002ik} matching procedure, is used together with {\sc Pythia8}~\cite{Sjostrand:2014zea} (version 8.223)
and {\sc Herwig7}~\cite{Bellm:2015jjp,Bellm:2017bvx} (version 7.1.2). For the latter, the default angular-ordered shower is employed. 
The same parton showers are employed for the LO results of {\sc Phantom}.
{\sc Pythia8} is also employed for the LO results of {\sc WHIZARD}. For the {\sc Powheg-Box}, the namesake
matching procedure is employed~\cite{Nason:2004rx,Frixione:2007vw}, together with {\sc Pythia8} (version 8.230). {\sc VBFNLO} serves as a matrix-element and phase-space provider
for the {\sc Matchbox} module~\cite{Platzer:2011bc} of {\sc
Herwig7}~\cite{Bellm:2015jjp,Bellm:2017bvx}, using an extended version of the Binoth Les Houches Accord
interface~\cite{Binoth:2010xt,Alioli:2013nda,Andersen:2014efa}. The {\sc Matchbox} module makes it
possible to choose between {\sc MC\-@NLO}-like and {\sc Powheg}-like
matching. As parton show\-er, both the default angular-ordered shower as
well as the dipole shower can be employed. We use here the subtractive (MC@NLO-type) matching to these parton-show\-er algorithms.
Whenever {\sc Pythia8}\ is used, the Monash tune~\cite{Skands:2014pea} is selected. Multiple-parton interactions are disabled.

Results are presented within the cuts described in Sec.~\ref{subsec:inputpar}, applied after shower and hadronisation (this implies that jets
are obtained by clustering stable hadrons, and not QCD partons). It follows that at the event-generation level, looser cuts (or even no cuts at all)
must be employed in order not to bias the results. This also implies that the tagging jets, whose momenta enter in the 
renormalisation and factorisation scales, Eq.~(\ref{eq:defscale}), are now defined without imposing
the $\Delta R_{\Pj\Pl}$ cut. The effect of this change is below one per cent at the level of the fiducial cross sections at NLO.

Compared to the fixed-order computations, a slightly different set-up has been employed for {\sc MG5\_aMC} in order to simplify the calculation: instead of generating the full
${\rm p}{\rm p}\to\mu^+\nu_\mu{\rm e}^+\nu_{\rm e}{\rm j}{\rm j}$ process, since it is dominated by doubly-resonant contribution, the
events are produced for the process with two stable ${\rm W}^+$-bosons (${\rm p}{\rm p}\to{\rm W^+}{\rm W^+}{\rm j}{\rm j}$), and the decay of these ${\rm W}^+$-bosons
is simulated with {\sc MadSpin}~\cite{Artoisenet:2012st} (ensuring spin correlations) before the parton shower. Since {\sc MadSpin}\ computes
the partial and total decay widths of the W bosons at LO accuracy only, while in Section~\ref{subsec:inputpar} the NLO width is employed,
an effect ($6\%$) on the normalisation is induced. 

\begin{table}
    \centering
    \begin{tabular}{c|l@{ $\pm$ }l}
      Code  &  \multicolumn{2}{c}{$\sigma[\rm{fb}]$}  \\
        \hline\hline
        {\sc MG5\_aMC}+{\sc Pythia8}&  $1.352 $ & $0.003$  \\
        {\sc MG5\_aMC}+{\sc Herwig7}&  $1.342 $ & $ 0.003$  \\
        {\sc MG5\_aMC}+{\sc Pythia8}, $\Gamma_{\rm resc}$&  $1.275$ & $0.003$  \\
        {\sc MG5\_aMC}+{\sc Herwig7}, $\Gamma_{\rm resc}$&  $1.266$ & $ 0.003$  \\
        {\sc Phantom}+{\sc Pythia8} &  $1.235 $ & $0.001$  \\
        {\sc Phantom}+{\sc Herwig7} &  $1.258 $ & $0.001$  \\
        {\sc VBFNLO}+{\sc Herwig7-Dipole} &  $1.3001$ & $0.0002$  \\
        {\sc WHIZARD}+{\sc Pythia8} &  $1.229$ & $0.001$  \\
    \end{tabular}
    \caption{\label{tab:PSratesLO} Cross sections at LO+PS accuracy.
    The {\sc MG5\_aMC} results with $\Gamma_{\rm resc}$
    are rescaled to account for the effect related to the W-boson width computed by {\sc MadSpin} (see the text for details).
    The uncertainties shown refer to estimated statistical errors of the Monte Carlo integrations.}
\end{table}

\begin{table}
    \centering
    \begin{tabular}{c|l@{ $\pm$ }l}
      Code  &  \multicolumn{2}{c}{$\sigma[\rm{fb}]$}  \\
        \hline\hline
        {\sc MG5\_aMC}+{\sc Pythia8}&  $1.491 ^{+1\%}_{-2\%} {}^{+2\%}_{-2\%} $ & $0.004$  \\
        {\sc MG5\_aMC}+{\sc Herwig7}&  $1.427 $ & $0.003$  \\
        {\sc MG5\_aMC}+{\sc Pythia8}, $\Gamma_{\rm resc}$&  $1.407$ & $0.003$  \\
        {\sc MG5\_aMC}+{\sc Herwig7}, $\Gamma_{\rm resc}$&  $1.346$ & $0.002$  \\
        {\sc Powheg-Box}+{\sc Pythia8}  & $1.3642$ & $0.0004$  \\
        {\sc VBFNLO}+{\sc Herwig7-Dipole} &  $1.3389 ^{+0\%}_{-1\%}$ & $0.0006$  \\
        {\sc VBFNLO}+{\sc Herwig7} &  $1.3067$ & $0.0006$  \\
    \end{tabular}
    \caption{\label{tab:PSratesNLO} Cross sections at NLO+PS accuracy.
    The {\sc MG5\_aMC} results with $\Gamma_{\rm resc}$
    are rescaled to account for the effect related to the W-boson width computed by {\sc MadSpin} (see the text for details). For
    {\sc VBFNLO}+{\sc Herwig7-Dipole}, the three-point scale uncertainties are shown, while for  {\sc MG5\_aMC}+{\sc Pythia8} the two displayed uncertainties
are respectively the nine-point scale uncertainty and the PDF one.
The uncertainties shown refer to estimated statistical errors of the Monte Carlo integrations.}
\end{table}

We now present the results of predictions matched to parton showers.
The total rates within VBS cuts are displayed in Tables~\ref{tab:PSratesLO} and
\ref{tab:PSratesNLO}, at LO and NLO
accuracy respectively. For {\sc MG5\_aMC},
the numbers with $\Gamma_{\rm resc}$ are rescaled to
take into account the width effects described in the above paragraph. At NLO accuracy, for {\sc MG5\_aMC} + {\sc Pythia8} and {\sc VBFNLO}+{\sc Herwig7-Dipole}, we also quote
theoretical uncertainties.
For the former, we show both PDF and scale uncertainties,\footnote{A preliminary study on PDF uncertainties in VBS has appeared
in Ref.~\cite{Dittmaier:2018zzz}.} obtained via exact reweighting~\cite{Frederix:2011ss} by varying independently the renormalisation and factorisation
scales by a factor of two around the central value, Eq.~(\ref{eq:defscale}) (nine-point variations).
For the latter, we show the
three-point scale uncertainties, obtained by considering correlated variations of the renormalisation, factorisation, and shower starting scale. Theory uncertainties should have very little dependence on the tool employed.
We observe that, once the width effect is taken into
account, total rates from different tools agree within some per cents ($\le7\%$), both at LO and NLO. Larger discrepancies, however, appear for differential observables, which we discuss in
the following. Theory uncertainties on the total rates are very small, regardless of whether scale variations are estimated with 
independent or correlated variations of the renormalisation and factorisation scales.
Concerning differential distributions, for each observable we display results in two plots, shown side-by-side. In the plot on the left (right), (N)LO+PS predictions are shown
with different colours in the main frame. In the inset, these predictions are compared in both cases with a fixed-order prediction at NLO accuracy (obtained with
{\sc VBFNLO}, \emph{i.e.}\ the VBS approximation with $s$-channel).
For the differential observables, the {\sc MG5\_aMC} predictions are \emph{not} rescaled to compensate for the width effect mentioned above. As for the table, we show theoretical uncertainties for the NLO+PS samples
obtained with {\sc VBFNLO} and {\sc MG5\_aMC}:
again, for the first the band corresponds to three-point variations, while for the second the darker (lighter) band corresponds to nine-point
scale variations (plus PDF uncertainties, linearly added).

\begin{figure*}
\centering
\includegraphics[width=0.47\textwidth]{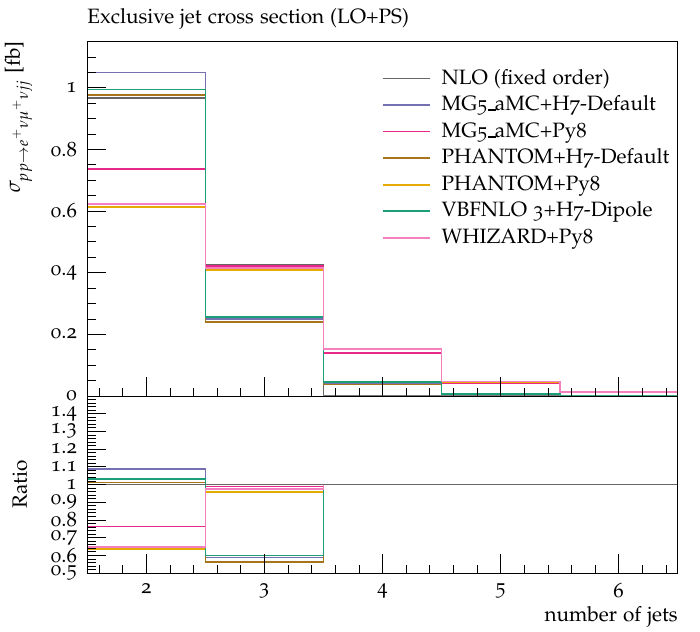}
\includegraphics[width=0.47\textwidth]{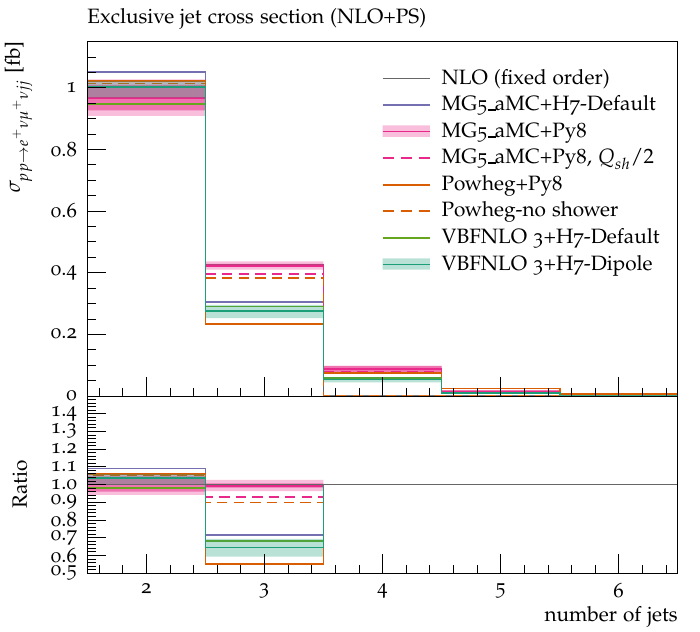}
\caption{Differential distribution in the
exclusive jet multiplicity
from predictions matched to parton showers, at LO (left) or NLO (right) accuracy (upper plot), compared with the fixed-NLO result computed with {\sc VBFNLO} (lower plot). At NLO+PS accuracy, for
    {\sc VBFNLO}+{\sc Herwig7-Dipole}, the three-point scale uncertainties are shown, while for {\sc MG5\_aMC}+{\sc Pythia8} the darker and lighter bands correspond
    respectively to the nine-point scale uncertainty and the scale and PDF uncertainties combined linearly.
    The predictions are obtained in the fiducial region described in Sec.~\protect\ref{subsec:inputpar}.}
\label{fig:PSnjet}
\end{figure*}

The first observable we investigate is the exclusive jet multiplicity, shown in Fig.~\ref{fig:PSnjet}. Looking at the LO+PS predictions, one can appreciate that the
main effects are driven by the parton shower that is employed ({\sc Herwig7} or {\sc Pythia8}), with the clear tendency of producing more radiation for the latter,
leading to higher jet multiplicities. Differences among tools that employ the same parton shower are typically smaller, and can be traced back to different values of the
initial scale of the parton shower (the {\tt scalup} entry of the Les Houches Event (LHE) file \cite{Alwall:2006yp,Butterworth:2010ym}). This event-by-event number corresponds
to the maximum hardness (translated into the shower-evolution variable) of the radiation that
can be generated by the shower.\footnote{At LO, the choice of such a scale is arbitrary and usually driven by common sense,
as it is the case for the factorisation and renormalisation scales.
    At NLO, one has the freedom to change the shower scale without losing formal NLO accuracy within the {\sc MC@NLO} matching,
    provided the Monte Carlo counterterms are also consistently
updated. In the {\sc Powheg} matching, the shower scale of the so-called $\tilde B$ events is fixed to the transverse momentum of the radiation
generated according the {\sc Powheg} Sudakov factor, while it can be changed in the remnant events.}
The main effect of NLO corrections for this (rather inclusive) observable is to stabilise the predictions for the two-jet bin, where discrepancies
among tools are reduced to about $10\%$. For the three-jet bin, which is described only at LO accuracy, differences among tools remain large, and are
possibly related to the underlying approximation performed ({\sc MG5\_aMC} is the only tool to use the full matrix element for the real radiation), in particular the inclusion of the $s$-channel contributions: the largest rate is predicted by
{\sc MG5\_aMC}, while the smallest rate is predicted by the {\sc Powheg-Box}, both matched to {\sc Pythia8}. Despite the fact that the same parton shower is employed, the way emissions are treated
is different among the two tools. In particular, for the {\sc Powheg-Box}, the first emission is generated with an internal Sudakov form factor (the
prediction dubbed {\sc Powheg-LHE} corresponds to stopping after the first emission), while for {\sc MG5\_aMC} there is an
interplay between the real-emission matrix element and the shower emission. For this observable we also show the prediction obtained
with {\sc MG5\_aMC+Pythia8} by reducing the shower-starting scale by a factor 2 with respect to the default value, dubbed 
{\sc MG5\_aMC+Py8, $Q_{sh}/2$}.\footnote{The reduction of the shower scale for {\sc MG5\_aMC+Herwig7} gives no visible effect on any of the observables 
discussed in this work.} The main effect of reducing the shower scale is that events migrate from the three-jet bin into the two-jet bin, \emph{i.e.}\ less radiation is generated. The
size of this effect on the jet rate is $+4\%$ ($-8\%$) on the two (three) jet bin, while the total rate within cuts is left unchanged.  

\begin{figure*}
\centering
\includegraphics[width=0.47\textwidth]{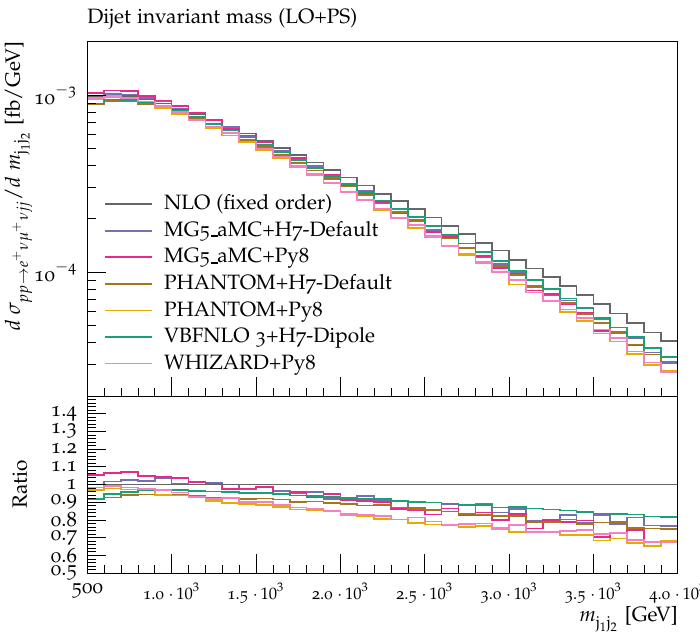}
\includegraphics[width=0.47\textwidth]{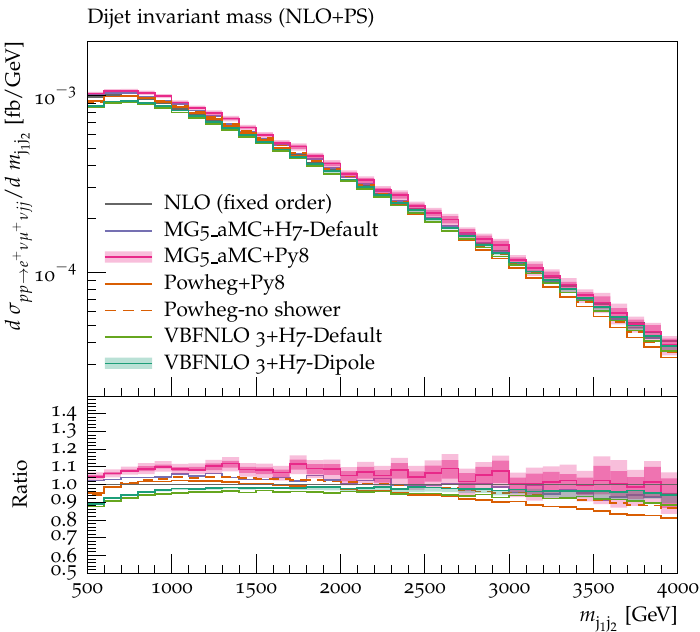}
\caption{Differential distribution in the
invariant mass of the two tagging jets
from predictions matched to parton showers, at LO (left) or NLO (right) accuracy (upper plot), compared with the fixed-NLO result computed with {\sc VBFNLO} (lower plot). At NLO+PS accuracy, for
    {\sc VBFNLO}+{\sc Herwig7-Dipole}, the three-point scale uncertainties are shown, while for {\sc MG5\_aMC}+{\sc Pythia8} the darker and lighter bands correspond
    respectively to the nine-point scale uncertainty and the scale and PDF uncertainties combined linearly.
    The predictions are obtained in the fiducial region described in Sec.~\protect\ref{subsec:inputpar}.}
\label{fig:PSmjj}
\end{figure*}

The next observable that we study is the invariant mass of the two tagging jets, shown in Fig.~\ref{fig:PSmjj}. For this observable, both at LO+PS and NLO+PS,
the spread of predictions matched with parton shower is rather small
($\lesssim 10\%$, if one compensates for the $6\%$ width effect for {\sc MG5\_aMC}).
The LO+PS predictions tend to be significantly softer than the fixed NLO one, with an effect of
about $-30\%$ at the end of the displayed range. At NLO+PS, this effect is mitigated, owing to the better description of the first QCD emission which is now driven by the real-emission matrix element.
For this observable (and all the others which are NLO accurate) the effect of reducing the shower scale is negligible, hence it is not shown.

\begin{figure*}
\centering
\includegraphics[width=0.47\textwidth]{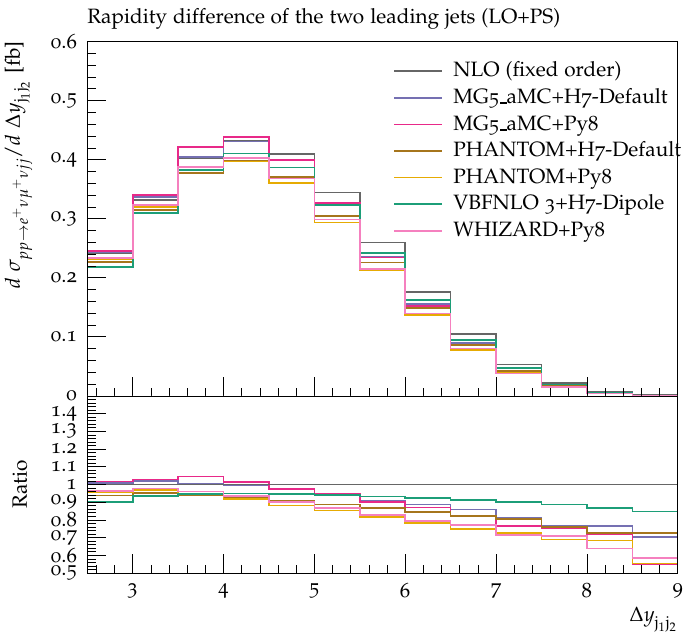}
\includegraphics[width=0.47\textwidth]{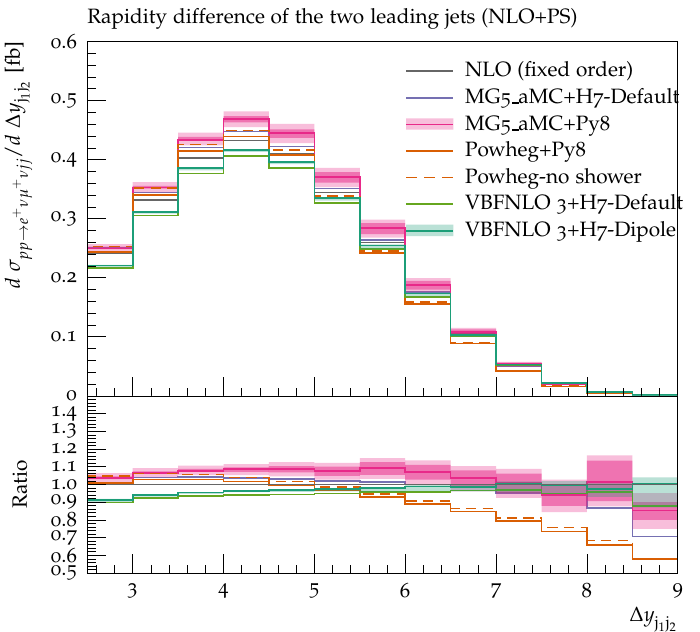}
\caption{Differential distribution in the
rapidity separation of the two tagging jets
from predictions matched to parton showers, at LO (left) or NLO (right) accuracy (upper plot), compared with the fixed-NLO result computed with {\sc VBFNLO} (lower plot). At NLO+PS accuracy, for
    {\sc VBFNLO}+{\sc Herwig7-Dipole}, the three-point scale uncertainties are shown, while for {\sc MG5\_aMC}+{\sc Pythia8} the darker and lighter bands correspond
    respectively to the nine-point scale uncertainty and the scale and PDF uncertainties combined linearly.
    The predictions are obtained in the fiducial region described in Sec.~\protect\ref{subsec:inputpar}.}
\label{fig:PSdyjj}
\end{figure*}

The rapidity difference between the two tagging jets, shown in Fig.~\ref{fig:PSdyjj}, has some similarities with the invariant-mass distribution.
At LO+PS all predictions show the tendency to deplete the large-separation region with respect to the fixed-order prediction, in a
quantitatively similar way,
except for {\sc VBFNLO+Herwig7} where the effect is mitigated. At 
NLO+PS, when the extra radiation is described by the real matrix element, such an effect is greatly reduced. A notable
exception is the {\sc Powheg-Box} prediction, which still shows a suppression at large separations.
Since such a suppression is already there for the {\sc Powheg-LHE} sample,
it is very likely that it is driven by the way the first emission is generated. A minor effect in the same direction is visible in the last two bins of the
{\sc MG5\_aMC+Herwig7} prediction (although with rather large statistical uncertainties).

\begin{figure*}
\centering
\includegraphics[width=0.47\textwidth]{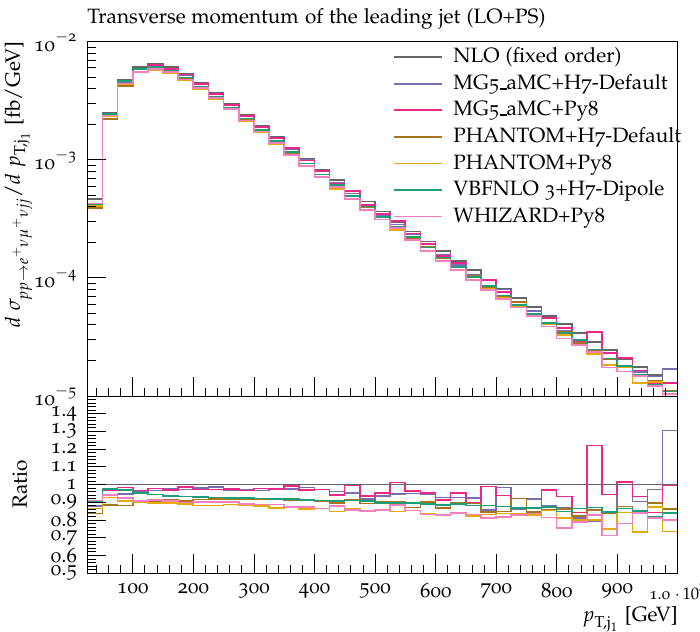}
\includegraphics[width=0.47\textwidth]{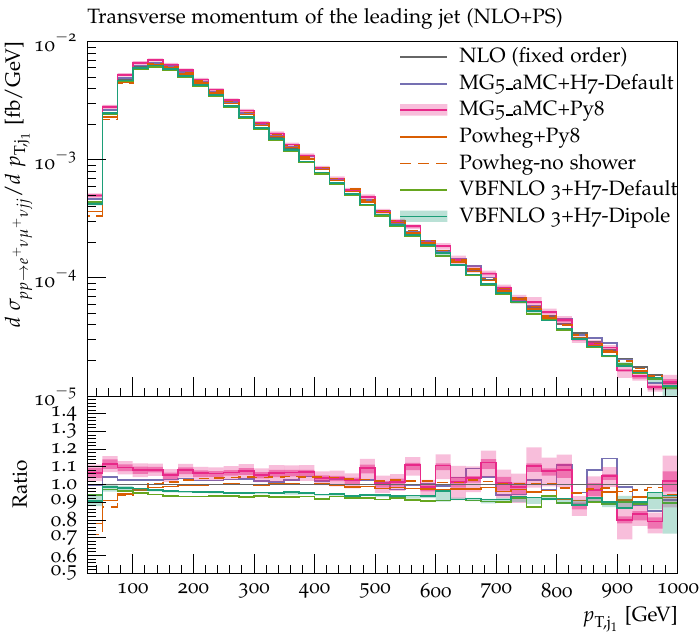}
\caption{Differential distribution in the
transverse momentum of the hardest jet
from predictions matched to parton showers, at LO (left) or NLO (right) accuracy (upper plot), compared with the fixed-NLO result computed with {\sc VBFNLO} (lower plot). At NLO+PS accuracy, for
    {\sc VBFNLO}+{\sc Herwig7-Dipole}, the three-point scale uncertainties are shown, while for {\sc MG5\_aMC}+{\sc Pythia8} the darker and lighter bands correspond
    respectively to the nine-point scale uncertainty and the scale and PDF uncertainties combined linearly.
    The predictions are obtained in the fiducial region described in Sec.~\protect\ref{subsec:inputpar}.}
\label{fig:PSpt1}
\end{figure*}

\begin{figure*}
\centering
\includegraphics[width=0.47\textwidth]{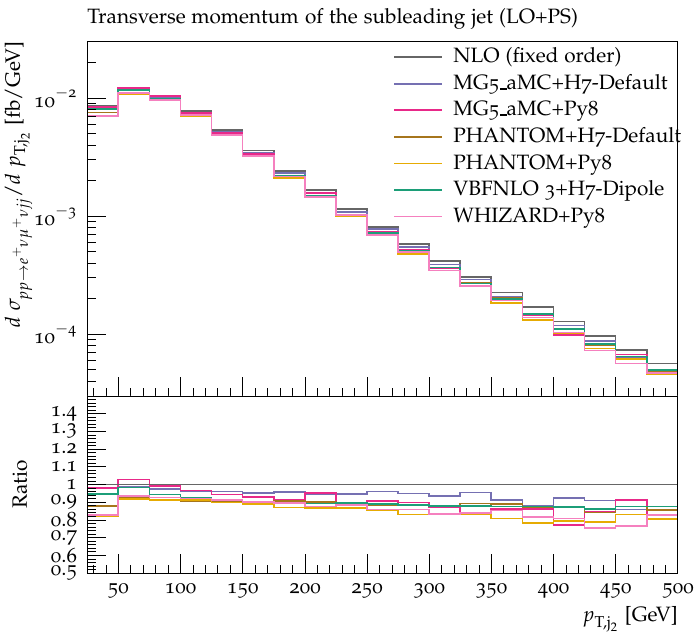}
\includegraphics[width=0.47\textwidth]{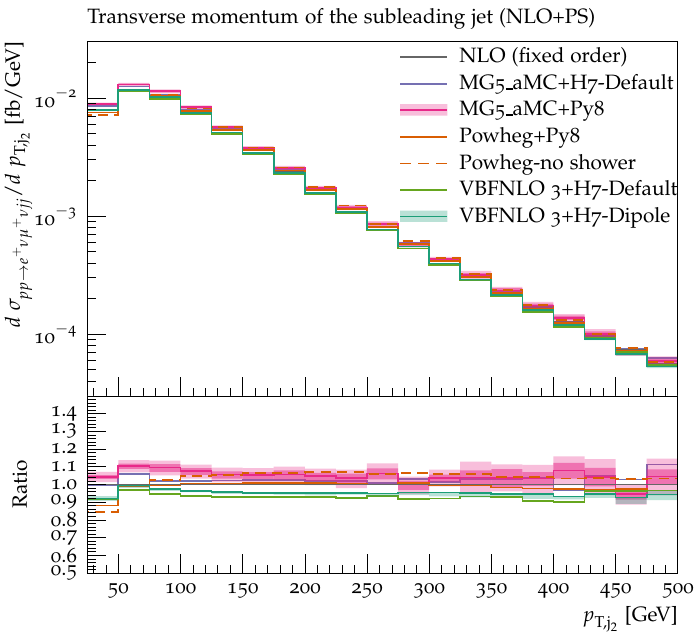}
\caption{Differential distribution in the
transverse momentum of the second-hardest jet
from predictions matched to parton showers, at LO (left) or NLO (right) accuracy (upper plot), compared with the fixed-NLO result computed with {\sc VBFNLO} (lower plot). At NLO+PS accuracy, for
    {\sc VBFNLO}+{\sc Herwig7-Dipole}, the three-point scale uncertainties are shown, while for {\sc MG5\_aMC}+{\sc Pythia8} the darker and lighter bands correspond
    respectively to the nine-point scale uncertainty and the scale and PDF uncertainties combined linearly.
    The predictions are obtained in the fiducial region described in Sec.~\protect\ref{subsec:inputpar}.}
\label{fig:PSpt2}
\end{figure*}

The transverse momentum of the hardest and second-hardest jets are shown in Figs.~\ref{fig:PSpt1} and~\ref{fig:PSpt2}, respectively. In general, for both observables,
predictions from different tools agree rather well with each other, with a spread at most at the 10\% level. At LO+PS, typically the transverse-momentum spectra are softer than
the fixed-NLO ones, and this effect is more marked for the second-hardest jet which, as expected, is more sensitive to the description of the extra radiation. Again, this
effect is mitigated by NLO corrections. The only feature that may be worth noticing among the NLO+PS predictions is the tendency of the {\sc Powheg-Box} to suppress the
hardest-jet spectrum at low transverse momentum ($\ptsub{\Pj_1}<100 \GeV$).

\begin{figure*}
\centering
\includegraphics[width=0.47\textwidth]{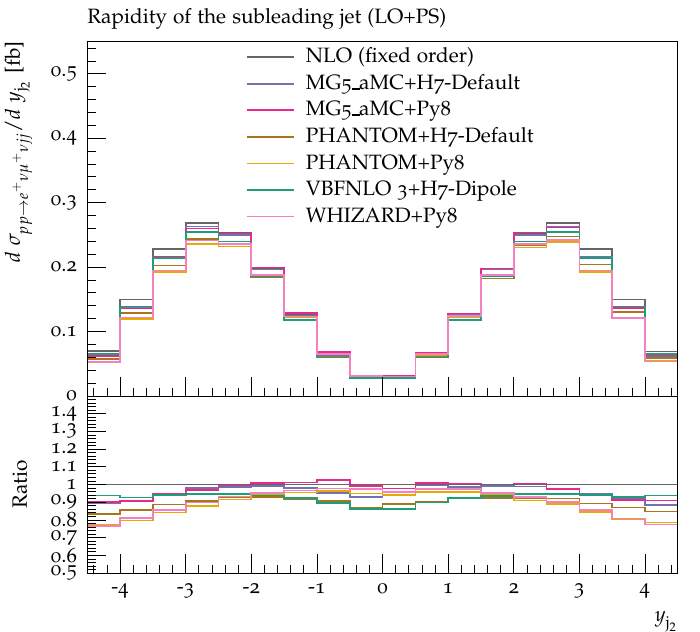}
\includegraphics[width=0.47\textwidth]{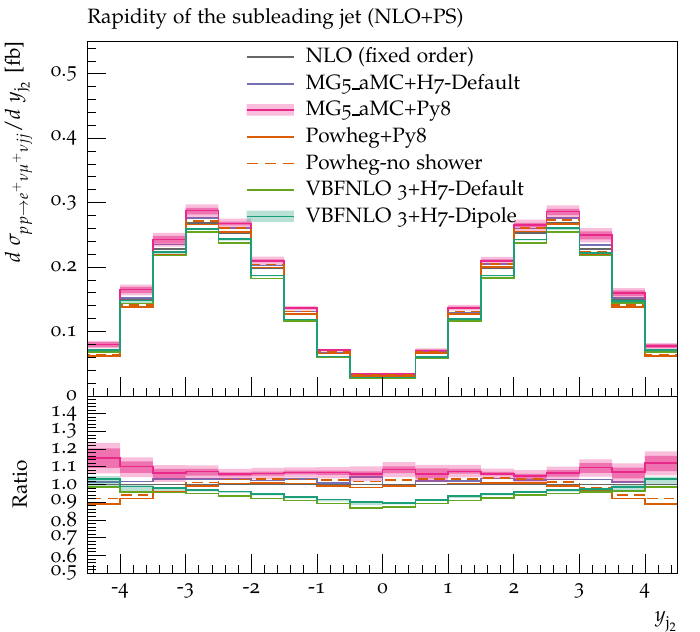}
\caption{Differential distribution in the
rapidity of the second-hardest jet
from predictions matched to parton showers, at LO (left) or NLO (right) accuracy (upper plot), compared with the fixed-NLO result computed with {\sc VBFNLO} (lower plot). At NLO+PS accuracy, for
    {\sc VBFNLO}+{\sc Herwig7-Dipole}, the three-point scale uncertainties are shown, while for {\sc MG5\_aMC}+{\sc Pythia8} the darker and lighter bands correspond
    respectively to the nine-point scale uncertainty and the scale and PDF uncertainties combined linearly.
    The predictions are obtained in the fiducial region described in Sec.~\ref{subsec:inputpar}.}
\label{fig:PSy2}
\end{figure*}

If we consider the rapidity of the second jet, Fig.~\ref{fig:PSy2}, we observe again rather small differences among tools, with the tendency towards a general
stabilisation at NLO+PS. However, some (small) differences in the shape remain at NLO+PS, which are worth to be briefly discussed: predictions
obtained with {\sc MG5\_aMC} are very close to the fixed-order prediction; the {\sc Powheg-Box}\ displays an enhancement of the central region, and a consequent suppression in the
peripheral region, while {\sc VBFNLO} shows an opposite behaviour. However, the effect is rather small, with the largest departure from the fixed-order prediction being
at most $10\%$.\footnote{If the setting {\tt SpaceShower:dipole\-Recoil = on} (discussed in the following)
is used when {\sc Pythia8}\ is employed together with the {\sc Powheg-Box}, the enhancement at central rapidities and the depletion
at small value of transverse momentum are partially compensated.}

\begin{figure*}
\centering
\includegraphics[width=0.47\textwidth]{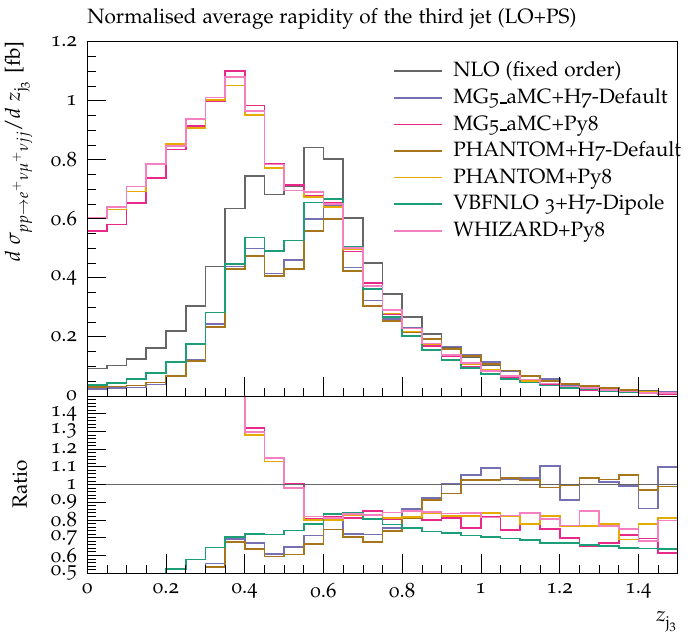}
\includegraphics[width=0.47\textwidth]{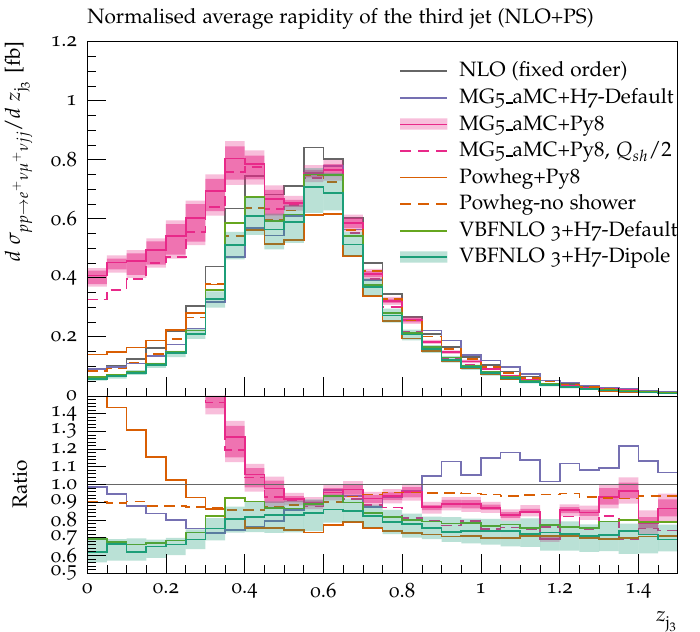}
\caption{Differential distribution in the
Zeppenfeld variable of the third-hardest jet
from predictions matched to parton showers, at LO (left) or NLO (right) accuracy (upper plot), compared with the fixed-NLO result computed with {\sc VBFNLO} (lower plot). At NLO+PS accuracy, for
    {\sc VBFNLO}+{\sc Herwig7-Dipole}, the three-point scale uncertainties are shown, while for {\sc MG5\_aMC}+{\sc Pythia8} the darker and lighter bands correspond
    respectively to the nine-point scale uncertainty and the scale and PDF uncertainties combined linearly.
    The predictions are obtained in the fiducial region described in Sec.~\protect\ref{subsec:inputpar}.}
\label{fig:PSz3}
\end{figure*}

Finally, focusing on the third jet, we conclude the list of differential observables by showing the Zeppenfeld variable defined in Eq.~\eqref{eq:Zeppenfeld}, Fig.~\ref{fig:PSz3}. This
variable is closely related to the third jet rapidity, and small (large) values of $z$ correspond to central (peripheral) rapidities. In general, for observables which involve the third jet, one
can clearly see a degradation of the agreement among the various tools, because of the poorer perturbative description of these observables. The Zeppenfeld variable is
a striking example: both at LO and NLO, the tendency of {\sc Pythia8} to generate more hard and central radiation, corresponding to low values of $z$,
is clearly visible. Such an effect, which is related to the way {\sc Pythia8} deals with the recoil of the radiation in VBF(VBS)-type processes,
can be mitigated by setting {\tt SpaceShower:dipole\-Recoil = on} in the {\sc Pythia8} input file.\footnote{This requires version $\ge8.230$.
Note that such a setting is not compatible with the NLO matching
in {\sc MG5\_aMC} (but it is compatible with the {\sc Powheg} matching). Also, this setting has other
effects, though smaller, on the rapidity spectra of the two hardest jets.} It is interesting to notice that
the effect survives beyond the first emission, as it can be observed by comparing {\sc Powheg-LHE} with {\sc Powheg+ Pythia8}, and that
it is only marginally attenuated when the shower scale is reduced. A similar
behaviour of {\sc Pythia8}
has also been observed in the study of EW production of a $\PZ$ boson in association with two jets (see the recent CMS measurement,
Ref.~\cite{Sirunyan:2017jej} Figure 12), where the experimental data seem to prefer the description by {\sc Herwig++}~\cite{Bahr:2008pv, Bellm:2013hwb}.
The central enhancement
is a bit mitigated if NLO+PS tools are used (compare LO+PS and NLO+PS from {\sc MG5\_aMC+Pythia8} with the fixed-NLO prediction), however even at NLO+PS the central region
($z_{j_3}<0.5$) is cursed by huge differences between tools. Large differences, reaching a factor 2, persist also away from the central region. These findings are consistent with behaviour displayed in Refs.~\cite{Jager:2011ms,Jager:2012xk,Jager:2013mu,Jager:2013iza,Schissler:2013nga} where the behaviour of NLO matching in VBS processes has been reported.\\

In conclusion, the comparison of tools including matching to parton shower clearly shows the benefits of the inclusion of NLO corrections: for most observables described
effectively at NLO accuracy differences between tools are at (or below) the $10\%$ level.
Some exceptions exist, \emph{e.g.}\ the rapidity separation of the two tagging jets, which on the one hand
clearly suggest not to rely on a single tool/parton shower, and on the other make it worth investigating more in detail the way QCD radiation is
generated, \emph{e.g.}\ when fully-differential computations at NNLO will become available (for VBF Higgs production, see Refs.~\cite{Cacciari:2015jma, Cruz-Martinez:2018rod}). It is a remarkable fact that, even for those observables that display small discrepancies,
the theoretical uncertainty obtained via scale variations (renormalisation, factorisation, and show\-er scale) 
systematically underestimates the spread of predictions. We note 
that in the only VBF process where NNLO corrections are known, \emph{i.e.}\ VBF Higgs production~\cite{Cacciari:2015jma, Cruz-Martinez:2018rod}, 
the NLO scale-uncertainty band does not include the NNLO prediction. This suggests that the NLO scale variation 
underestimates the size of the perturbative uncertainty. Again, this stresses the need
 to employ at least two different tools in order to obtain a more realistic estimate of theoretical uncertainties. Finally, the size of discrepancies for observables that are described at a lower perturbative accuracy, notably those related to the third jet, suggests that
experimental analyses should rely as little as possible on those observables and, in any case, use conservative estimates of the theory
uncertainties. On the one hand, in order to improve the
description of these observables, a simulation of VBS+j at NLO accuracy, currently unavailable but within the reach of modern
automated tools, is certainly desirable.
On the other hand, measurements of processes with similar
colour flow (EW production of a single vector boson plus jets,
VBF, \ldots) can certainly help in order to discriminate which tools perform better in the comparison with data~\cite{Aaboud:2017emo,Sirunyan:2017jej}.

\section{Conclusions and recommendations}
\label{sec:conclusion}

In the present article, a detailed study of the process $\Pp\Pp\to\mu^+\nu_\mu{\rm e}^+\nu_{\rm e}\,\Pj\Pj+\mathrm{X}$ at the LHC has been presented, 
mainly focused on the EW production mechanism which involves the scattering of massive vector bosons.
Until very recently, when the complete calculation became available for the NLO QCD corrections (order $\mathcal O (\alphas\alpha^6)$),
the so-called VBS approximation was the standard for this kind of simulations. For this reason, various theoretical predictions 
have been compared to the full computation, both in a typical VBS fiducial region and also in more inclusive phase space.
We have precisely quantified the differences that arise for several physical observables, 
in particular for the di-jet invariant mass and the rapidity separation of the leading two jets.
This is the first time that such an in-depth study is performed.
Besides the study of fixed-order predictions, we have also investigated the impact of parton showering.
To that end, several LO and NLO event generators 
which are able to perform matching to parton showers have been employed, and various observables have been thoroughly compared.
While in general observables which are described at NLO accuracy show reasonable agreement among the tools, larger differences can 
appear for those observables described at a lower accuracy, such as those that involve the third jet.
In particular such differences are quite prominent in the central-rapidity region, and are the largest for those simulations which employ {\sc Pythia8}.
The effect
has been understood, and it can be partially mitigated by changing the recoil scheme of {\sc Pythia8} to distribute momenta within initial--final colour connections. These findings make it worth 
to further investigate these issues not only in the theoretical community, but also by experimental collaborations, for example by 
measuring related observables for similar processes.

The last part of our work is devoted to 
 remarks and recommendations concerning the usage of theoretical predictions by experimental collaborations.
\begin{itemize}
    \item As found in Ref.~\cite{Biedermann:2017bss}, the NLO EW corrections of order $\mathcal{O}{\left(\alpha^{7}\right)}$ are 
        the dominant NLO contribution to the process $\Pp\Pp\to\mu^+\nu_\mu{\rm e}^+\nu_{\rm e}\,\Pj\Pj+\mathrm{X}$.
        It is thus highly desirable to combine them with NLO-QCD predictions matched with parton shower, or at least to include them into experimental analyses. 
        Since, as shown in Ref.~\cite{Biedermann:2016yds}, 
        these large EW corrections originate from the Sudakov logarithms which factorise, we recommend to combine them with QCD 
        corrections in a multiplicative way. The estimate of missing higher-order EW corrections can be obtained, 
        in a first approximation, by considering $\pm \delta^2_{\rm NLO EW}$,\footnote{The quantity $\delta_{\rm NLO EW}$ is defined through the relation $\sigma_{\rm NLO EW} = \sigma_{\rm LO} \left(1+ \delta_{\rm NLO EW}\right).$} while the missing higher-order mixed QCD-EW corrections 
        can be estimated by taking the difference between the multiplicative and additive prescriptions.
        For more detailed studies of the combination of QCD and EW higher-order corrections, see
        \emph{e.g.}\, Ref.~\cite{Czakon:2017wor} in the context of top-pair production, or Ref.~\cite{Lindert:2017olm} for SM 
        backgrounds for dark matter searches at the LHC.
        
    \item For the typical fiducial region used by experimental collaborations for their measurements, 
        the agreement between the approximations and the full calculation is satisfactory given 
        the current experimental precision, as well as the one foreseen for the near future \cite{CMSCollaboration:2015zni,CMS:2016rcn,ATL-PHYS-PUB-2017-023}.
        Nonetheless, care has to be taken when using such approximations, in particular if more inclusive phase-space cuts are used.

    \item In addition to the standard interpretation of EW signal versus QCD background, 
        combined measurements should also be presented as they are better defined theoretically. In fact, while at LO 
        the interference term can be included in the background component, at NLO the separation of EW and QCD components becomes more blurred, as, \emph{e.g.}\,
        at the order $\mathcal{O}{\left(\alphas\alpha^{6}\right)}$ both types of amplitudes contribute.
        Therefore, a combined measurement including the EW, QCD, and interference contributions is desirable.
        Note that with such a measurement a comparison to the SM would
        be straightforward and still be sensitive to the EW component.
        In addition, the QCD component could be subtracted based on a
        well-defined Monte Carlo prediction.

    \item Since the inclusion of NLO QCD corrections gives a better control of extra QCD radiation and reduces the ambiguities related to the 
        matching details and/or the parton shower employed, we encourage the use of NLO-accurate event generators in experimental analyses. In doing
        so, special care should be employed in order to estimate the theoretical uncertainties, as the standard prescription based on 
        renormalisation and factorisation-scale variation is clearly inadequate. Rather, different combinations of generators and parton showers should be employed.

    \item The present study has focused on the orders $\mathcal{O}{\left(\alpha^{6}\right)}$ at 
    LO and $\mathcal{O}{\left(\alphas\alpha^{6}\right)}$ at NLO. NLO computations and publicly-available tools also exist for the QCD-induced process~\cite{Rauch:2016pai,Melia:2010bm,Melia:2011gk,Campanario:2013gea,Baglio:2014uba,Biedermann:2017bss,Alwall:2014hca}.

    \item For practical reasons, we have focused on the ${\rm W}^+{\rm W}^+$ signature. Nonetheless, 
    the observed features 
    (\emph{e.g.}\ validity of the VBS approximation or comparison of theoretical predictions matched to parton shower) should 
    be qualitatively similar for other VBS signatures with massive gauge bosons. For these other signatures, similar quantitative studies should be performed.
\end{itemize}

\section*{Acknowledgements}

The authors are grateful to the members of the VBSCan collaboration for several discussions and comments on this work. The author also thank
the {\sc Pythia8} developers, in particular Stefan Prestel, Torbj{\"o}rn Sj{\"o}strand, and Peter Skands for 
discussions and clarifications about the third-jet rapidity spectrum. MZ would like to thank Andreas Papaefstathiou, Christian Reuschle and David Grellscheid 
for their help with {\sc Herwig7}. RF, HSS, and MZ thank the {\sc MadGraph5\_\-aMC@NLO} authors for discussions, and
they are particularly grateful to Valentin Hirschi for comments on this manuscript.

The authors would like to acknowledge the contribution of the COST Action CA16108 which initiated this work.
Moreover, this work was supported by several STSM Grants from the COST Action CA16108.
Many authors acknowledge hospitality from Nikhef, where part of this work has been performed.

BB, AD, and MP acknowledge financial support by the
German Federal Ministry for Education and Research (BMBF) under
contract no.~05H15WWCA1 and the German Research Foundation (DFG) under
reference number DE 623/6-1.
SD and CS acknowledge support by the state of Baden-W\"urttemberg through bwHPC and the DFG through grant no. INST 39/963-1 FUGG and grant DI 784/3.
The work of BJ was supported in part by the Institutional Strategy of the University of T\"ubingen (DFG, ZUK 63) and in part by
the BMBF under contract number 05H2015.
AK acknowledges financial support by the Swiss National Science Foundation (SNF) under contract 200020-175595.
MR acknowledges funding from the European Union's Horizon 2020 research and innovation programme as part of the Marie Sk\l{}odowska-Curie Innovative Training Network MCnetITN3 (grant agreement no. 722104).
HSS is supported by the ILP Labex (ANR-11-IDEX-0004-02, ANR-10-LABX-63).
GZ is supported by ERC Consolidator Grant HICCUP (no. 614577).
MZ is supported by the Netherlands National Organisation for Scientific Research (NWO).

\bibliographystyle{utphys}
\bibliography{article}

\end{document}